%% file: main.tex
\def\shownotes{0}
\def\revision{1}
\documentclass[sigconf,balance=false,dvipsnames]{acmart}
\usepackage{popets}
\setcopyright{popets}
\copyrightyear{YYYY}

\acmYear{YYYY}
\acmVolume{YYYY}
\acmNumber{X}
\acmDOI{XXXXXXX.XXXXXXX}
\acmISBN{}
\acmConference{Proceedings on Privacy Enhancing Technologies}
\input{macros}

\begin{document}

\title[Gaze3P: Gaze-Based Prediction of User-Perceived Privacy]{Gaze3P: Gaze-Based Prediction of User-Perceived Privacy}

\author{Mayar Elfares}
\affiliation{
  \institution{University of Stuttgart, Germany}
  \country{}
}
\email{mayar.elfares@vis.uni-stuttgart.de}

\author{Pascal Reisert}
\affiliation{
  \institution{University of Stuttgart, Germany}
  \country{}
  }
\email{pascal.reisert@sec.uni-stuttgart.de}

\author{Ralf Küsters}
\affiliation{
  \institution{University of Stuttgart, Germany}
  \country{}
  }
\email{ralf.kuesters@sec.uni-stuttgart.de}

\author{Andreas Bulling}
\affiliation{%
 \institution{University of Stuttgart, Germany}
  \country{}
  }
\email{andreas.bulling@vis.uni-stuttgart.de}

\renewcommand{\shortauthors}{Elfares et al.}
%\pr{Page limit:12 pages without reference/appendix}
\input{sections/abstract}




\received{20 February 2007}
\received[revised]{12 March 2009}
\received[accepted]{5 June 2009}

\begin{teaserfigure}
\centering
  \includegraphics[width=0.8\columnwidth]{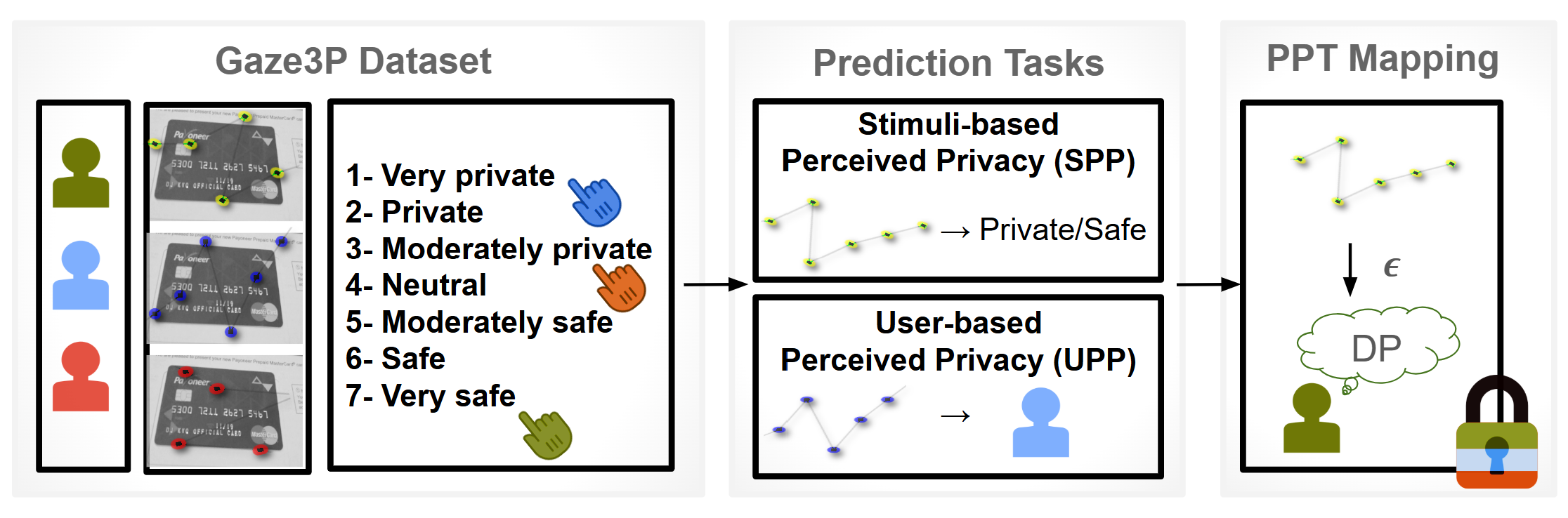}
  \caption{Users deal with various types of information in daily life that can have vastly different privacy requirements, e.g., personal photographs, passwords, or social media posts.
  %During everyday interactions, users are exposed to various stimuli (e.g. personal photographs or social media ads) with different privacy levels.
  \methodName~is the first large-scale dataset that allows for the systematic study of user-perceived privacy. We report extensive experiments demonstrating the feasibility of predicting perceived privacy from human eye gaze. We also show how predicted privacy can be used to optimise the parameters of privacy-preserving techniques for data analysis and learning, such as Differential Privacy (DP), to better align them with user expectations.
  %\andreas{might be good to show actual stimulus images (like in fig 2) instead of the pictograms/icons. We could then even remove fig 2. Also, the figure illustrates how users look at these images but not the gaze-based prediction part. I would also not show eyes etc in light grey as these are actually important. Maybe an entirely new teaser could be interesting to discuss: we have three entities: gaze on stimuli/images, user-perceived privacy (in the form of ratings/labels), the prediction tasks, and the parameter optimisation. maybe a figure showing three "circles/clouds", one for each. in the middle a Gaze3P label. And then arrows between them (e.g. prediction, analysis/understanding, etc) in the same colour as gaze3P illustrating that the dataset enables them. Essentially, the teaser should/could illustrate the three contributions (analysis/undestanding, prediction, optimisation)\mayar{updated}}
  }
  %\Description{...}
  \label{fig:teaser}
\end{teaserfigure}

%prediction/mapping, arrows...

\maketitle

%\linenumbers

% Let's mark changes with the corresponding package. Acronyms are \added[id=mayar]{mayar},\replaced[id=andreas]{andreas}{RK},\deleted[id=PR]{PR}.

\section{Introduction}
\label{sec:intro}
\input{sections/intro}

\section{Preliminaries} \label{sec:preliminaries}
\input{sections/preliminaries}

\section{Related Work} \label{sec:related_work}
\input{sections/related_work}

\section{\methodName~Dataset} \label{sec:dataset}
\input{sections/dataset}

\section{\methodName~Tasks} \label{sec:benchmarks}
\input{sections/benchmarks}

\section{Privacy-Preserving Applications}
\label{sec:applications}
\input{sections/applications}

% \section{Complementary Tasks} \label{sec:complementary}
% \input{sections/complementary}

\section{Discussion}
\label{sec:discussion}
\input{sections/discussion}

\section{Conclusion}
\label{sec:conclusion}
\input{sections/conclusion}

\begin{acks}
We thank Manpa Barman and Yanhong Xu for their assistance in recruiting participants. M. Elfares was funded by the Ministry of Science, Research and the Arts Baden-Württemberg in the Artificial Intelligence Software Academy (AISA). 
%P. Reisert and R. Küsters were supported by the German Federal Ministry of Education and Research under Grant Agreement No. 16KIS1441 (CRYPTECS project) and the German Research Foundation under Grant No. 548713845.\pr{This probably should change to a newer project}
\end{acks}
\nolinenumbers
\bibliographystyle{ACM-Reference-Format}
\bibliography{references}
\linenumbers
\appendix
\section*{Appendix}
\label{sec:appendix}
\input{sections/appendix}

\end{document}

%% file: macros.tex
\newcommand{\methodName}{Gaze3P}
\ifnum\shownotes=1
    \newcommand\mayar[1]{\textcolor{blue}{Mayar: #1}}
    \newcommand\pr[1]{\textcolor{cyan}{Pascal: #1}}
    \newcommand\andreas[1]{\textcolor{orange}{Andreas: #1}}
    \newcommand\rk[1]{\textcolor{red}{Ralf: #1}}
\else 
    \newcommand\mayar[1]{}
    \newcommand\pr[1]{}
    \newcommand\andreas[1]{}
    \newcommand\rk[1]{}
\fi
\usepackage[switch,mathlines]{lineno}
\usepackage{enumitem}
\usepackage{tabularx}
\usepackage{supertabular}
\usepackage{makecell}
\usepackage{multirow}
\usepackage{soul}
\usepackage[capitalise]{cleveref}
\renewcommand{\ref}[1]{\cref{#1}}
\crefname{footnote}{footnote}{footnotes}
\newcommand{\bpara}[1]{\bigskip\noindent\emph{\bf #1.}~}
\usepackage[group-separator={,},group-minimum-digits={3}]{siunitx}
\usepackage[edges]{forest}
\usepackage{tikz,hyperref}
\usetikzlibrary{arrows.meta}

\ifnum\revision=1
    \ifnum\shownotes=1
        \usepackage[commandnameprefix=ifneeded]{changes}
    \else
        \usepackage[commandnameprefix=ifneeded,authormarkup=none]{changes}
    \fi
    \setaddedmarkup{\textcolor{Black}{#1}}
    \setdeletedmarkup{\textcolor{Red}{\st{#1}}}
\fi
\ifnum\revision=0
    \usepackage[final,commandnameprefix=ifneeded]{changes}
\fi

    \definechangesauthor[name={PR}]{PR}
    \definechangesauthor[name={mayar}]{mayar}
    \definechangesauthor[name={andreas}]{andreas}
    \definechangesauthor[name={RK}]{RK}
\usepackage{framed}
\usepackage{quoting}

 \colorlet{shadecolor}{GreenYellow}
\usepackage{lipsum}
\newenvironment{shadedquotation}
 {\begin{shaded*}
  \quoting[leftmargin=0pt, vskip=0pt]
 }
 {\endquoting
 \end{shaded*}
}

%% file: sections/abstract.tex
\begin{abstract}
Privacy is a highly subjective concept and perceived variably by different individuals.
Previous research on quantifying user-perceived privacy has primarily relied on questionnaires.
Furthermore, applying user-perceived privacy to optimise the parameters of privacy-preserving techniques (PPT) remains insufficiently explored.
To address these limitations, we introduce \methodName~-- the first dataset specifically designed to facilitate systematic investigations into user-perceived privacy.
Our dataset comprises gaze data from 100 participants and 1,000 stimuli, encompassing a range of private and safe attributes.
With \methodName~we train a machine learning model to implicitly and dynamically predict perceived privacy from human eye gaze.
Through comprehensive experiments, we show that the resulting models achieve high accuracy.
Finally, we illustrate how predicted privacy can be used to optimise the parameters of differentially private mechanisms, thereby enhancing their alignment with user expectations.
\end{abstract}

%% file: sections/intro.tex
Privacy, particularly as it is perceived by individuals, is a complex and deeply subjective construct that varies significantly across contexts, cultures, and personal experiences \cite{meier2024privacy,dinev2006extended,culnan1999information}.
Unlike technical privacy, which can be quantified through cryptographic guarantees or formal metrics, perceived privacy refers to an individual's internal judgment about the sensitivity or appropriateness of data sharing \cite{beldad2011trust, kokolakis2017privacy}. 
Understanding and quantifying user-perceived privacy is essential because it directly influences users’ willingness to engage with digital systems, share information, or consent to data sharing requests \cite{johnson1974privacy, meier2024privacy,boenisch2022individualized}. Therefore, accurately quantifying perceived privacy helps designers create user-aligned privacy mechanisms, improve transparency, and ultimately enhance user satisfaction and system usability \cite{hsu2014differential, lee2011much, jorgensen2015conservative, boenisch2022individualized}.

The ability to quantify user-perceived privacy levels also has significant potential for optimising the parameters of security protocols, such as Differential Privacy (DP) \cite{Dwo06}.
However, despite continuing discussions, the problem of how to map users' privacy perception to protocol parameters remains unsolved \cite{culnan1999information, dinev2006extended, meier2024privacy, knijnenburg2022user, ostendorf2020neglecting, culnan1999information, dinev2006extended}.
A key reason for this failure is the large number of factors that affect privacy perception, such as (i) oversight of the situational diversity \cite{masur2018situational}, (ii) neglect of within- vs. between-subject variations \cite{meier2024privacy}, (iii) effects of biases, heuristics, or impulsivity on online user behaviour \cite{knijnenburg2022user, ostendorf2020neglecting}, and, most importantly, (iv) the scarcity of the available behavioural data that encapsulates all relevant aspects.
Previous work mainly relied on explicit feedback, such as questionnaires, which has been shown to not align well with users' satisfaction \cite{orekondy2017advisor, steil2019privaceye}, especially since user judgement dynamically changes depending on context, behaviour, or knowledge \cite{hoyle2020privacy}.

In this work, we explore a novel approach: The use of \emph{human eye gaze as an implicit and dynamic source of information on user-perceived privacy}.
\mayar{cite implicit and dynamic}\added{In particular, our approach does not require a user to explicitly provide input or make active decisions, e,g, clicking a button or selecting a privacy setting. 
Naturally, the feedback derived from gaze is responsive to changing contexts, such as different content, tasks, or user states, i.e. as the user interacts with new stimuli, their gaze behaviour adapts, and our approach therefore continuously updates itself accordingly.}
Prior research has shown that eye gaze contains rich information about the user, such as identifiers \cite{cantoni2015gant}, quasi-identifiers \cite{sammaknejad2017gender}, confidential attributes (e.g. user activities \cite{steil2015discovery}, attentive \cite{faber2018automated, vertegaal2003attentive} and cognitive states \cite{huang2016stress,bulling11_ubicomp}, or information about private situations \cite{steil2019privaceye}). 
Consequently, we try to answer our main research question RQ1:
\begin{shadedquotation}
\noindent\textit{RQ1:} \textbf{Can human eye gaze be used as an implicit and dynamic indicator of user‑perceived privacy?}
\end{shadedquotation}

To this end, we present \methodName~ \added[id=mayar]{(\textbf{Gaze}-based \textbf{P}rediction of user-\textbf{P}erceived \textbf{P}rivacy) }~-- the first large-scale dataset for studying user-perceived privacy from the perspective of human eye gaze.
\methodName~includes gaze data (i.e. where, when, and how a person looks) of $100$ participants viewing $\num{1000}$ natural images showing different objects, some with private attributes (e.g. credit cards and medical history) as shown in \cref{fig:sample}.
The dataset also provides user ratings of perceived privacy on a scale from 1 (very private) to 7 (very safe) for each image.
The full dataset, including all annotations, will be made publicly available upon acceptance (cf. \cite{code} for the implementation).
Using our new dataset, we then explore: %\linelabel{q2:2}
\begin{shadedquotation}
\noindent\textit{RQ2:} \textbf{How accurately can ML models predict privacy perceptions solely from gaze?}
\end{shadedquotation}
We therefore present different tasks, focusing on the automatic prediction of users' perceived privacy solely from gaze behaviour.
\added[id=mayar]{These tasks correspond to privacy-related problems or objectives that an algorithm is trained to address using our collected data. Once trained, the algorithm can be used to generalise its learned solution to apply to the same class of problems for previously unseen individuals.}
\added[id=mayar]{Since it is not feasible to a priori determine which features can be reliably extracted from gaze data—nor whether potentially confounding factors can be effectively disentangled—we employed machine learning algorithms to automatically identify privacy-related patterns associated with each task.}
These tasks include: \textbf{Stimuli-based Perceived Privacy (SPP)} tasks to infer how private a stimulus (e.g. image) is and \textbf{User-based Perceived Privacy (UPP)} tasks to infer information about the user (e.g. privacy expertise or identity).
%s: \rk{add questions for these terms as well}Privacy Expertise Prediction, User Privacy Profiling, and Privacy-aware Gaze Identification.
Our ML models demonstrate that human eye gaze provides accurate predictions of perceived privacy.

\added[id=mayar]{Hence, we explore a third research question:}
\begin{shadedquotation}
\noindent\textit{RQ3:} \textbf{Can gaze‑based predictions of perceived privacy be integrated into privacy-preserving frameworks (such as Differential Privacy) to optimize utility while aligning with user expectations?}
\end{shadedquotation}
We use the \methodName~predictions of user-perceived privacy to optimise the parameters of differentially private mechanisms.
Differentially private mechanisms obfuscate sensitive data samples such that only a limited amount of information about the private data can still be deduced from the obfuscated output of the mechanism. 
The exact amount of acceptable leakage depends on a privacy budget parameter $\varepsilon>0$, which determines the obfuscating noise added by the mechanism. 
If $\varepsilon$ is small, the privacy guarantee becomes stronger, but usually, the output of the mechanism is less accurate, and usability decreases.
It is therefore important not to choose the privacy budget $\varepsilon$ too small, i.e., to only add the minimal amount of noise that guarantees a target privacy level.
The optimisation of DP-parameters has therefore seen much attention in recent years \added{\cite{jorgensen2015conservative, boenisch2022individualized, niu2021adapdp}}.

Our new dataset \methodName~and the resulting ML model~predictions provide a new way to determine $\varepsilon$, which reflects a user's perceived privacy.
Depending on the actual use case, we propose different mappings from perceived privacy levels to $\varepsilon$-values. 
We evaluate how each mapping affects the utility of the obfuscated output dataset and show that our gaze-based approach outperforms previous work.

\bpara{Contributions} In summary, our work makes the following contributions: 
\begin{enumerate}[leftmargin = *]
    \item We present \methodName~-- the first large-scale dataset for studying user-perceived privacy using human eye gaze, \added[id=mayar]{providing a dynamic and implicit user feedback}.    
    \item We propose several novel learning tasks focusing on predicting user-perceived privacy from human eye gaze.
    These tasks cover different aspects of privacy and also allow us
    to explore potential applications and limitations of gaze-based privacy perception.    
    \item We demonstrate how gaze-based predictions can be used to optimise parameters of privacy-preserving techniques. Specifically, we introduce a novel approach that maps predicted privacy levels to DP's privacy parameter $\varepsilon$ and show that aligning DP with user expectations improves the data utility in data analysis and learning. %
\end{enumerate} 

%% file: sections/preliminaries.tex
\bpara{Eye Tracking}
Gaze data is typically collected using eye-tracking devices that record the position and movement of a user's eyes relative to a visual stimulus or screen. Modern eye trackers employ infrared light to detect corneal reflection and pupil centre, enabling accurate estimation of gaze coordinates at high temporal resolutions. The raw gaze signal is then processed into interpretable features such as fixations, saccades, and pupil dilation:
\begin{itemize}[leftmargin=*]
\item \emph{Fixations} refer to time periods where the eye remains focused on a specific location, typically lasting 100–400 ms. They are indicative of visual attention and cognitive processing of that region. 

\item \emph{Saccades} are rapid eye movements between fixations used to reposition the fovea to new visual targets, lasting 20–80 ms. These movements are ballistic, and their patterns can inform about scanning behaviour and search strategies.

\item \emph{Pupil dilation} is a physiological response modulated, amongst others, by both environmental lighting and cognitive load. Increased dilation was linked to heightened mental effort, emotional arousal, or attentional demand.
\end{itemize}

Together, these gaze features provide a rich, temporally fine-grained source of implicit user feedback (i.e. without requiring direct input or explicit interaction).
We refer the reader to \cite{hessels2025fundamentals, hooge2025fundamentals, nystrom2025fundamentals, niehorster2025fundamentals} for details about eye tracking and gaze behaviour analysis.

\added[id=mayar]{By transforming the continuous, high-dimensional gaze signals into quantitative features—such as fixations, saccades, and pupil diameters—machine learning (ML) models are provided with informative inputs that capture essential characteristics of user interaction or cognitive states. These features are then passed into ML algorithms, enabling the models to learn underlying statistical patterns or associations within the data. This learned structure allows the models to perform specific tasks such as classification, regression, or clustering. Ultimately, this feature-to-model pipeline allows ML systems to generalize from training data and make accurate, data-driven inferences about new, unseen inputs.}

\bpara{Differential Privacy (DP)} DP is a mathematical framework that ensures privacy by limiting the impact of any single data point on the output of a computation.
A randomised algorithm $M$ satisfies $\varepsilon$-DP if, for all datasets $D$ and $D'$ differing by at most one element, and for all measurable subsets $S$ of the output space:
\begin{align}\label{eq:dp}
    \Pr[M(D) \in S] \leq e^{\varepsilon} \cdot \Pr[M(D') \in S],
\end{align}
where $\varepsilon\geq 0$ is the privacy budget, controlling the privacy loss.
A small $\varepsilon$ means that $D$ and $D'$ are (almost) not distinguishable given a set of outputs $S$.
From an adversarial perspective, an adversary $\mathcal A$ challenged to distinguish $D$ and $D'$ given an output set $S$ will output the dataset which is more likely, e.g. $D$ if \added[id=PR]{$\Pr(D|M(D)\in S)\geq \frac{1}{2}\geq \Pr(D|M(D')\in S)$.}
In the most extreme case of \cref{eq:dp} we have $\Pr(M(D)\in S)=e^{\varepsilon}\Pr(M(D')\in S)$ and hence \added[id=PR]{$\Pr(D|M(D)\in S)=e^{\varepsilon}\Pr(D'|M(D)\in S)=e^{\varepsilon}(1-\Pr(D|M(D)\in S))\Rightarrow \Pr(D|M(D)\in S)=\frac{e^{\varepsilon}}{1+e^{\varepsilon}}$}. Thus, the (absolute) advantage of an adversary is bounded by $\operatorname{adv}_{\mathcal A}\leq 2\frac{e^{\varepsilon}}{1+e^{\varepsilon}}-1=\frac{e^{\varepsilon}-1}{e^{\varepsilon}+1}$ (cf. \cref{sec:appendix-dp} for more details).

%% file: sections/related_work.tex
\bpara{User Perceived Privacy}
As more data is being collected, shared, and processed, sensitive insights about the user's personality, intentions, and preferences are being leaked \cite{meier2024privacy, kairouz2021advances}.
Hence, to better protect user privacy, prior works have investigated the psychological mechanisms of privacy decision-making \cite{meier2024privacy}, self-disclosure \cite{dinev2006extended}, and the related cost-benefit analysis \cite{culnan1999information}.
They showed that the user-perceived privacy dynamically changes according to the user's context, behaviour, and knowledge.
Other works focused on privacy-in-context (i.e. contextual integrity) \cite{nissenbaum2009privacy,  wang2016examining} and further showed that user-perceived privacy is affected by culture, activities (e.g., online shopping vs. online banking), and platforms (desktop vs. mobile).
Prior works \cite{steil2019privaceye, liu2019differential, kairouz2021advances} often used generic privacy mechanisms that remain static throughout the interaction.
These mechanisms are typically predefined at the onset of a session (e.g., at the initiation of a protocol) and fail to account for the dynamic nature and context-dependent fluctuations of the users' privacy judgments \cite{hoyle2020privacy}. 

\bpara{Eye Gaze and Privacy Perception}
The results of the aforementioned works usually rely on user questionnaires. 
However, when using questionnaires, users often fail to follow their own privacy preferences \cite{orekondy2017advisor, plutzer2019privacy, gilbert17measuring}.
In this paper, we propose to use human eye gaze (instead of questionnaires) to implicitly capture the dynamics of user-specific privacy perception.
Eye-tracking data is already widely used to study human behaviour and cognition \cite{eckstein2017beyond, carter2020best, hessels2025fundamentals, hooge2025fundamentals, nystrom2025fundamentals, niehorster2025fundamentals}.
\added[id=mayar]{Prior research has demonstrated that gaze patterns can reflect cognitive processes such as risk perception and habituation \cite{anderson2016your, guerra2023seeing, guerra2024visual}. For instance, studies have shown that increased risk is often associated with longer fixations and heightened visual scanning, indicating deeper cognitive engagement \cite{toma2023gazing, mormann2016role, harrison2019eye}. Conversely, habituation to repeated stimuli can lead to reduced gaze variability and decreased attention, even in the presence of sensitive information—a challenge for maintaining consistent privacy awareness \cite{anderson2016your}. \methodName~builds on these insights and} hypothesises that gaze can also be used as an indicator of user-perceived privacy.
The idea to use gaze to detect privacy-sensitive situations is not completely new and has been explored in \cite{steil2019privaceye}, where the users' eye movements and first-person video were recorded using an egocentric (head-mounted) camera. 
However, \cite{steil2019privaceye} focuses on detecting privacy-sensitive situations rather than quantifying privacy perception. 
It also only features a small set of 17 participants in a free-viewing task and relies on recording and processing the scene imagery, which might break privacy 
\cite{orekondy2017advisor}.
In this paper, we instead focus on the implicit and dynamic privacy perception feedback solely through gaze.

\bpara{Other Privacy-Related Gaze Applications}
\added[id=mayar]{Prior works at the intersection of eye tracking and privacy mainly focus on (i) eye-based authentication \cite{shiqing20oculock, liebers2020gaze}, (ii)  privacy considerations and guidelines \cite{gressel23_pcm, katsini20role}, (iii) secure AR/VR applications \cite{shiqing20oculock, li2021kaleido, bozkir2023eyetracked, david2021privacy, david2021towards, david2022your}, (iv) UI design for secure interactions \cite{katsini20role}, (v) secure gaze data sharing \cite{elfares2022federated, privateeyes, elfares2025qualiteye, bozkir2021diffrential, david2021privacy, david2021towards, david2022your}, and (vii) information leakage and attacks on gaze data \cite{wang24gazeploit, yimin18eyetell, privateeyes, sonnichsenattentionleak}. 
Apart from the different focus, \methodName, also differs in (i) the purpose of gaze, i.e., some prior work \cite{wang24gazeploit, shiqing20oculock, li2021kaleido} uses gaze for explicit interaction (e.g., authentication \cite{shiqing20oculock}
or gaze rays in VR \cite{li2021kaleido}) while \methodName~ uses gaze as an implicit (passive) signal to infer perceived privacy, (ii) unlike prior work, we focus on the cognitive aspect of gaze in privacy, beyond the gaze location estimation \cite{wang24gazeploit, yimin18eyetell}, (iii) privacy positioning, i.e. \methodName~is user-centric and proactive (helping users protect their own privacy through implicit gaze behaviour) while prior works either focus on system-enforced protection for eye tracking \cite{elfares2022federated, privateeyes, elfares2025qualiteye, li2021kaleido, david2021privacy, david2021towards, david2022your}, user authentication \cite{shiqing20oculock, liebers2020gaze}, or demonstrating privacy vulnerabilities \cite{katsini20role, sonnichsenattentionleak}.}

\bpara{Personalised Differential Privacy}
In classical differential privacy (DP), the privacy parameter $\varepsilon$ is chosen independently of the subjective privacy perception of an individual user but instead provides the same privacy guarantees to all users \cite{jorgensen2015conservative}.
This is despite the fact that users have varying expectations about acceptable privacy levels.
As a result, a certain DP-privacy level $\varepsilon$ might not offer enough protection for some users while over-protecting others \cite{jorgensen2015conservative, galli2022machine}.
Personalised Differential Privacy (PDP) \cite{ebadi2015differential, acharya2024personalized, jorgensen2015conservative, alaggan2015heterogeneous, boenisch2022individualized, cummings2015accuracy, niu2021adapdp} is an extension of standard DP that introduces flexibility by tailoring privacy protections based on individual preferences, allowing for a more nuanced balance between privacy and utility.
Alaggan et al. \cite{alaggan2015heterogeneous} first introduced the theoretical concept of PDP through linear pre-processing (a.k.a stretching) of the input data.
In their approach, the input data is scaled according to each individual's privacy preference before applying a differentially private mechanism.
Cummings and Durfee \cite{cummings2020individual} generalised the PDP framework to a broader class of mechanisms.
They proposed a constructive method for implementing personalised privacy guarantees directly within the mechanism design.
However, they demonstrated that computing an optimal personalised mechanism under these conditions is NP-hard.
Later works \cite{fallah2022bridging, cummings2023advancing} used weighted moment estimation of each data point according to the privacy level.
In these approaches, the privacy level specified by each user was used to assign a weight to their corresponding data contribution during the statistical estimation process.
This weighting strategy ensured that data from users requiring stronger privacy protections (i.e., lower $\varepsilon$ values) exert less influence on the aggregate statistics, while users with looser privacy requirements (higher $\varepsilon$ values) contribute more significantly.
Other works \cite{li2017achieving, boenisch2023have} proposed partitioning the data into separate groups and then assigning different privacy levels.
Jorgensen et al. \cite{jorgensen2015conservative}, followed by Niu et al. \cite{niu2021adapdp} and Ebadi et al. \cite{ebadi2015differential}, relied on excluding or sub-sampling some data samples according to their privacy levels.
While all of these methods were designed for data analysis, Boesnisch et al. designed a privacy-preserving training mechanism for machine learning models that integrated individual privacy levels directly into the optimisation process \cite{boenisch2023have, boenisch2022individualized}.
They adapted the gradient computation and noise addition to reflect user-specific privacy budgets.

Similarly, we determine a suitable privacy budget $\varepsilon$ based on behavioural data rather than assigning it arbitrarily or using assumptions.
Our work is the first to do so using human gaze data.
This approach ensures that the privacy budgets more accurately reflect users' actual perceptions and expectations of privacy.
Grounding $\varepsilon$ in observed user behaviour not only improves the practical relevance and usability of privacy-preserving systems but also helps bridge the gap between formal privacy theory and human-centred privacy concerns, ultimately leading to more trustworthy and adaptive data handling practices.

%% file: sections/dataset.tex
%\rk{In introductory sentence would be nice here....What are we going to do}\andreas{you could first point out that there is a lack of data to study these questions. Either too small or insuitable annotations etc. This serves as a motivation to record our own large-scale dataset. What were the objectives when collecting it? what guided the decisions?}\pr{Just a subsitute, pleas fill in something meaningful:}
Due to the lack of available datasets, in this section, we present our new gaze-based dataset \methodName. 
Our large-scale data collection is essential for deriving statistically significant insights into users' perceptions of privacy and for empirically validating the use of gaze as a reliable indicator of perceived privacy.
By offering a standardized dataset, \methodName~aims to facilitate reproducible research and enable the development and evaluation of models that infer perceived privacy dynamically and without explicit feedback through gaze patterns.
% We use \methodName~in \cref{sec:benchmarks} to train a machine learning model that can predict preceived privacy from gaze data.

\bpara{Eye tracker} 
For gaze data collection, we used an Eyelink eye tracker that provides binocular gaze data at a sampling rate of \SI{2}{kHz}.
As is common practice in laboratory eye tracking studies, we used a chin rest to stabilise participants' heads.
Images were shown on a computer screen with a resolution of 1920x1080 pixels and a size of \SI{545}{mm} x \SI{303}{mm}.
The eye-to-screen distance was \SI{700}{mm}.
The proportion of the calibrated area was set to 0.63 x 0.88 to stay within the trackable range of the system with an HV13-13 (horizontal/vertical 13 targets) calibration type for better spatial accuracy across the entire screen. 
The recorded gaze data was then processed into fixations, saccades, and pupil information (cf. \cref{sec:preliminaries}).
%and depicts where, when, and how a participant looks at the screen, in terms of screen coordinates and time on the corresponding visual stimuli.
More details and visualisations can be found in \cref{apx:eye}.

\begin{figure}[t]
\centering
  \includegraphics[width=0.5\textwidth]{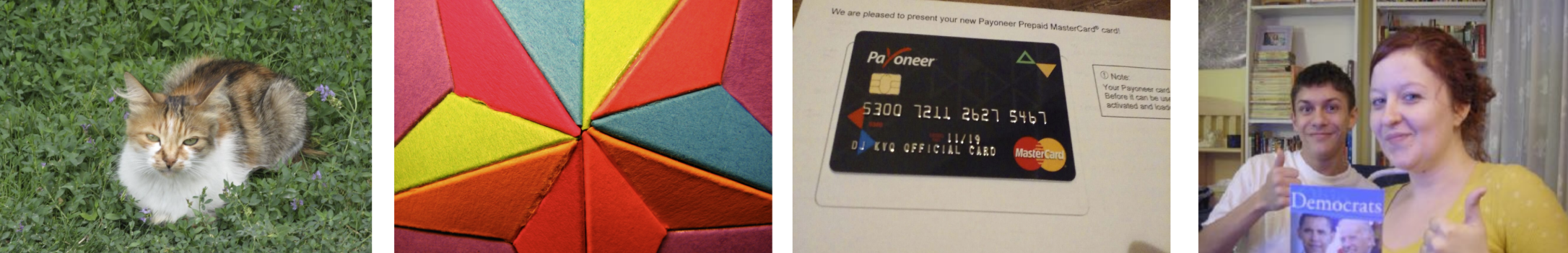}
  \caption{Sample images from the VISPR dataset with safe (e.g. cat, colours) and private (e.g. credit card, political opinion) attributes}
  \label{fig:sample}
\end{figure}
\bpara{Stimuli}
%\linelabel{q9:1}
\added[id=mayar]{We randomly sampled a subset of images from the VISPR dataset \cite{orekondy2017advisor} as our stimuli since it is the only publicly-available dataset that contains privacy-related attributes. 
See \cref{fig:sample} for sample images.}
The dataset contains 68 attributes categorised into nine attribute groups. 
The attributes were compiled according to the guidelines for the EU Data Protection Directive (GDPR) 95/46/EC, the US Privacy Act of 1974, and the data sharing rules in online social networks. %\linelabel{q9:2}
\added[id=mayar]{It further includes reliable and consistent attribute annotations by letting multiple annotators follow detailed labelling guidelines.}
\added[id=mayar]{We ensured a balanced distribution across the annotated privacy attributes. This stratified sampling approach was employed to maximise attribute coverage and mitigate sampling bias, given the practical constraint on the number of stimuli each participant could reasonably view during the experiment. Despite this limitation, the resulting dataset remains relatively large and representative, supporting robust analysis of visual privacy perception across a diverse range of attributes.}
Our dataset includes 1,000 images with corresponding private and safe (i.e. non-private) attributes. 
The attribute categories in VISPR \cite{orekondy2017advisor} %\linelabel{q20-a:1} 
are:
\begin{itemize}[leftmargin=*]
    \item \textbf{Personal information:} e.g. age, gender, fingerprint, signature.
    \item \textbf{Documents:} e.g. credit card, passport, national ID.
    \item \textbf{Medical:} e.g. medical history, hospital tickets, physical disability.
    \item \textbf{Employment:} e.g. occupation, work occasion.
    \item \textbf{Life:} e.g. culture, religion, political opinion, sexual orientation.
    \item \textbf{Relationship:} e.g. personal, social, professional.
    \item \textbf{Whereabouts:} e.g. landmark, home address.
    \item \textbf{Online activity:} e.g. date/time of activity, username, password.
    \item \textbf{Automobile:} e.g. license plate, vehicle ownership.
\end{itemize}

\bpara{Participants}
We initially recruited $103$ participants through university mailing lists and notice boards.
We had to exclude three participants due to calibration failures \cite{krafka2016eye}.
This resulted in a final group of $100$ participants ($37$ females and $63$ males).
Participants were aged between $18$  and $35$ years, had different nationalities ($25$) and different (self-reported) privacy knowledge ($12$ experts)\footnote{\label{footnote:msc}\added[id=mayar]{All experts are MSc or PhD holders of information security degrees.}}. All participants had normal or corrected-to-normal vision. We refer to \cref{apx:eye} for more demographic details.

\bpara{Experiment design}
After arriving in the lab, participants were first informed about the general purpose and procedures of the study.
Following \cite{sattar2020deep}, we explicitly asked participants to consider the data as their own, e.g. their phone gallery.
%Participants were first shown a standard example task  with $3$ images to familiarize them with the setup and reduce learning effects, e.g. initial biases and cognitive load.
The experiment consisted of four blocks with 25 participants each.
Each block included two tasks: free-viewing and a search task. In the former, participants were shown a stimulus (e.g. an image of a credit card) for five seconds and asked to rate its privacy-sensitivity for sharing on a scale from 1 (very private) to 7 (very safe) following \cite{steil2019privaceye}.
Similarly, for the search task, participants were asked to search for a specific image (by a category, e.g. \textit{document}) in a 2 x 2 image collage, and click the mouse and rate its sensitivity once found, following \cite{sattar2020deep}.
The same attributes were presented in both tasks, but each task featured different sets of images within those attributes.
Each task included three practice trials and $50$ recorded trials with randomized order of (non)private stimuli (\SI{50}{\percent}-\SI{50}{\percent} ratio). 
The resulting dataset includes a set of triplets of \{stimulus, gaze patterns (timestamped coordinates), and privacy rating\} for each participant in each task. 
Refer to \cref{apx:eye} for more details on the dataset structure and data collection software.

%\rk{It doesn't become clear what exact data is collected here. Somehow Gaze data (i.e., coordinates of where people looked/gaze points/paths) plus what privacy-level they chose?}\mayar{added.}

% \pr{Maybe repeat how the final dataset looks like. E.g. tuples (pic id, user id, ground truth, scan path,...)?}\mayar{Added to appendix}

%\rk{I would expect to see a description of the kind of data that was actually recorded, and hence, what can actually be found in the data set. I.e., some extended description of what is mentioned in the intro. So a paragraph starting with: The Data Set. ....}
%\pr{As mentioned before, a description of the format of the dataset could be helpful for the reader.}

\bpara{Compliance with the privacy and ethics guidelines}
The data was collected and processed according to the standards, guidelines, \added[id=mayar]{and approval} of the ethical committee of the authors' institution, with participants' consent, remuneration, and pseudo-anonymisation procedures.
In particular, the privacy and ethical guidelines of the Menlo Report \cite{2012-dittrich-mraf} were satisfied.
%\pr{Explicitly mentioned in the call.}

%% file: sections/benchmarks.tex
Our Gaze3P dataset enables new analyses that shed light on how users cognitively and behaviourally respond to privacy-relevant stimuli.
\added[id=mayar]{Given the difficulty in determining beforehand which features can be meaningfully extracted from gaze data and whether confounding variables can be effectively separated, we adopted a machine learning-based approach to automatically uncover privacy-related patterns corresponding to each task. This method enables us to objectively evaluate our hypothesis: that gaze behaviour may function as a proxy for perceived privacy.}

More specifically, in this section, we explore how well our gaze-based dataset \methodName~is suited for two groups of learning tasks: %\rk{you should make clear here what the difference between these terms is? Something like: SPP is about this and UPP is about that....}
(i) \emph{stimuli-based perceived privacy} (SPP) tasks that infer how private a stimulus (e.g. image) is in \cref{sub:spp} and (ii) \emph{user-based perceived privacy (UPP)} tasks that infer information about the user (e.g. privacy expertise or identity) in \cref{sub:upp}.
Each task entails learning a distinct mapping or pattern within the gaze data, such as predicting privacy ratings from gaze, classifying stimuli as private or non-private, or identifying user-specific privacy preferences based on demographic or behavioural features. 
\added[id=mayar]{For implementation, we used a basic Scikit-learn framework with default parameter settings. This provided a standardised and unbiased baseline for model training, avoiding manual feature selection or tuning that could skew results or introduce overfitting.}
%We remark that we train all our models on annotated gaze data (i.e. with user-specified privacy ratings as a ground-truth to enable evaluation of ML models), such that they generalise to unseen users or stimuli later without the users explicitly providing their ratings.
\added[id=mayar]{It is important to note that all models in this study are trained on annotated gaze data, where user-specified privacy ratings serve as ground-truth labels to facilitate supervised learning and enable robust evaluation of model performance. The purpose of this training procedure is to allow the model to learn association patterns for each task.
During testing and deployment, however, only the raw gaze data is provided as input to the model—without any accompanying user-specified ratings. This setup reflects a practical application scenario, wherein the model is expected to generalise to previously unseen users or stimuli and autonomously infer privacy-related judgments based solely on gaze behaviour.}
This not only facilitates the development of scalable and user-adaptive privacy-aware systems but also provides empirical insights into the underlying mechanisms of perceived privacy.

In the following, we describe the two training tasks in detail and evaluate the performance of our trained models w.r.t standard baselines, namely decision tree (DT), support vector machine (SVM), logistic regression (LR), random forest (RF), K-nearest neighbour (KNN), and transformer (TF) models (cf. \cref{tab:spp}).
To study individual stimuli, we focused on the free-viewing data of our dataset.
%\rk{Here or at the end of the section or maybe every subsection, we should summarize the main take away.}

\subsection{Stimuli-based Perceived Privacy (SPP)}\label{sub:spp}
Quantifying perceived privacy levels helps understand how users feel about their privacy protections. Privacy perception is mainly influenced by the nature of the stimuli, e.g. their type (e.g. images) and the content being shared or observed (e.g. a credit card). 
A quantitative relation between the stimuli and the corresponding privacy perception can help in designing more context-aware and effective privacy-preserving mechanisms.
We, therefore, propose four main SPP tasks:

\bpara{(1) Binary Privacy Perception} \label{task:detection}
Given solely the gaze data, the binary privacy perception task \cite{steil2019privaceye} aims to determine whether a user is exposed to or interacts with potentially sensitive information.
In addition to the pioneering work of Steil et al. \cite{steil2019privaceye}, we also want to determine how the different setups affect the binary perception, e.g. how the perception of a specific user (intra-user setting) varies for different stimuli in comparison to how the perception of many users (inter-user setting) varies.

\bpara{(2) Privacy Level Perception} 
\label{task:quantification}
This task aims to map the gaze data as inputs to a privacy level as output. The dataset includes a ground truth of 7 different privacy levels, following Steil et al. \cite{steil2019privaceye}, indicated by participants for each stimulus. Note that, unlike Steil et al. \cite{steil2019privaceye}, we process all classes instead of combining them into two.

\bpara{(3) Contextual Privacy Perception}
\label{task:contextual}
Integrating contextual information such as demographics or user expertise can potentially improve predictions of user-perceived privacy levels\footnote{Other contextual information like the type of application, time of the day, and the type of platform, has been shown to influence the user privacy perception too (cf. \cref{sec:related_work}). To simplify the setup, we did not include this information in our dataset; however, our approach naturally extends to more detailed datasets.}
Contextual information is provided as additional features to the model, capturing the user's age, gender, nationality, and privacy expertise.

\bpara{(4) Private Attribute Recognition}
\label{task:attribute}
Prior works \cite{cantoni2015gant, sammaknejad2017gender}
showed that gaze is a good predictor of the user's private attributes, such as age and gender. Other works \cite{orekondy2017advisor, steil2019privaceye} recognise the private attribute directly from images as inputs to their models. Here, we focus on predicting what private attribute the user is looking at, given the gaze data alone. The task becomes more challenging as semantics become more complex and diverse.

\bpara{SPP applications}
%\linelabel{q15:1} 
Before we discuss our results on tasks (1)--(4), we want to outline how stimuli-based perceived privacy results can be used in applications. For example, the quantified values in Tasks 1-3 (classified per attribute in Task 4, if needed) can be mapped to the corresponding parameters in privacy-preserving protocols. Such parameters can include $\varepsilon$ values in DP \cite{dwork2014algorithmic, Dwo06}, model update perturbation or gradient clipping thresholds in federated learning \cite{mcmahan2017communication, kairouz2021advances}, and similar hand-picked parameters in K-anonymity \cite{samarati1998protecting}, L-diversity \cite{machanavajjhala2007diversity}, T-closeness \cite{li2006t}, privacy auctions \cite{ghosh2011selling, zhang2020selling}, synthetic data generation \cite{liu2022privacy}, etc. 
Other stimuli-specific insights are important for multiple applications, such as access control models (to define who can access which piece of information, e.g. attribute-based access control (ABAC) \cite{hu2015attribute} which uses attributes for dynamic access control and human-in-the-loop privacy controls (e.g. Instagram’s 'Restrict' feature \cite{parmelee2020insta} for controlling interactions, e-mail spam filters that allow manual corrections \cite{androutsopoulos2004learning}, and marketing campaigns that respect the user's perceived privacy preferences \cite{martin2017role}).
We refer to \cref{sec:applications} for sample applications \added[id=mayar]{and to \cref{apx:apps} for further details.}

\bpara{SPP Baselines and Evaluation}
As baselines, we ran the aforementioned basic models.
We evaluate the accuracy of the models with cross-validation and test sets to ensure that the model generalises well to unseen data. \added[id=mayar]{Implementation details can be found in \cref{apx:implementation}.}
We also perform statistical hypothesis testing on the extracted eye-tracking features to examine whether the observed differences across experimental conditions or participant groups are statistically significant. These tests assess whether variations in metrics such as fixation duration, saccade amplitude, or pupil dilation are likely to reflect systematic effects rather than random noise or individual variability. To quantify the strength of evidence against the null hypothesis (i.e., that there is no meaningful difference between groups or conditions), we compute $p$-values. These values represent the probability of observing the given data, or something more extreme, under the assumption that the null hypothesis is true. A $p$-value below a conventional threshold (typically 0.05) is considered indicative of a statistically significant effect.
Hence, we made the following key observations in relation to SPP Tasks 1-4:

\begin{table}
  \caption{Accuracy of the SPP tasks \cref{sub:spp}.}
  \label{tab:spp}
  \begin{tabular}{lccc}
    \toprule
      &  \textbf{\makecell{Binary Privacy \\ Perception}} & \textbf{\makecell{Privacy Level\\ Perception}} & \textbf{\makecell{Contextual Privacy \\ Perception}}\\
    \midrule
    \multicolumn{4}{c}{\textbf{Person-independent}}\\
     DT & 0.54 & 0.19 & 0.23 \\
    SVM & 0.64 & 0.34 & 0.37  \\
    LR & 0.63 & 0.34 & 0.36\\
    RF & 0.60 & 0.29 &  0.32\\
    KNN & 0.56 & 0.21 & 0.24\\
    TF & 0.52 & 0.34 & 0.37\\
    \midrule
    \multicolumn{4}{c}{\textbf{Person-specific}} \\
     DT & 0.61 & 0.23 & 0.24 \\
     SVM & 0.70 & 0.38 &  0.39\\
     LR & 0.75 & 0.40  & 0.40\\
     RF & 0.66 & 0.35 & 0.37\\
     KNN & 0.62 & 0.30 & 0.31\\
     TF & 0.81 & 0.52 & 0.54\\
    \midrule
    \multicolumn{4}{c}{\textbf{Down-sampling}}\\
     DT & 0.48 & 0.15 & 0.16 \\
     SVM & \textbf{0.57} & 0.30 & 0.33  \\
     LR & 0.55 & 0.28 & 0.29 \\
     RF & 0.52 & 0.24 & 0.26 \\
     KNN & 0.45 & 0.20 & 0.20 \\
     TF & 0.46 & 0.30 & 0.31 \\
    \midrule
    \multicolumn{4}{c}{\textbf{Gaze + Stimuli (ours)}}\\
     DT  & 0.89 & 0.73 &  0.72 \\
     SVM & 0.93 & 0.81 & 0.83\\
     LR & 0.92 & 0.80 & 0.81\\
     RF & 0.90 & 0.82 &  0.80 \\
     KNN & 0.88 & 0.79 & 0.82\\
     TF & 0.97 & 0.87 & 0.90\\
    \midrule
    \multicolumn{4}{c}{\textbf{Gaze + Stimuli (PrivacEye)}}\\
     DT &0.58 &  - & - \\
     SVM &  0.67 &  - & - \\
     LR & 0.66 & - & - \\
     RF & 0.70 & - & - \\
     KNN & 0.58 & - & - \\
     TF & 0.75 & - & - \\
     \midrule
     \multicolumn{4}{c}{\textbf{Gaze Saliency Maps}}\\
     DT & 0.45 & 0.17 & 0.21 \\
    SVM & 0.52 & 0.30 & 0.31  \\
    LR & 0.57 & 0.29 & 0.30\\
    RF & 0.52 & 0.23 &  0.27\\
    KNN & 0.50 & 0.18 & 0.18\\
    TF & 0.46 & 0.29 & 0.32\\
     \midrule
     \multicolumn{4}{c}{\textbf{ Residual Gaze Maps}}\\
     DT & 0.23 & 0.11 & 0.13 \\
    SVM & 0.21 & 0.15 & 0.22  \\
    LR & 0.19 & 0.17 & 0.08\\
    RF & 0.27 & 0.19 &  0.21\\
    KNN & 0.09 & 0.10 & 0.23\\
    TF & 0.31 & 0.18 & 0.19\\
  \bottomrule
\end{tabular}
\end{table}

\subsubsection*{Inter- and intra-user variations} 
\begin{figure}
\centering
  \includegraphics[width=0.8\columnwidth]{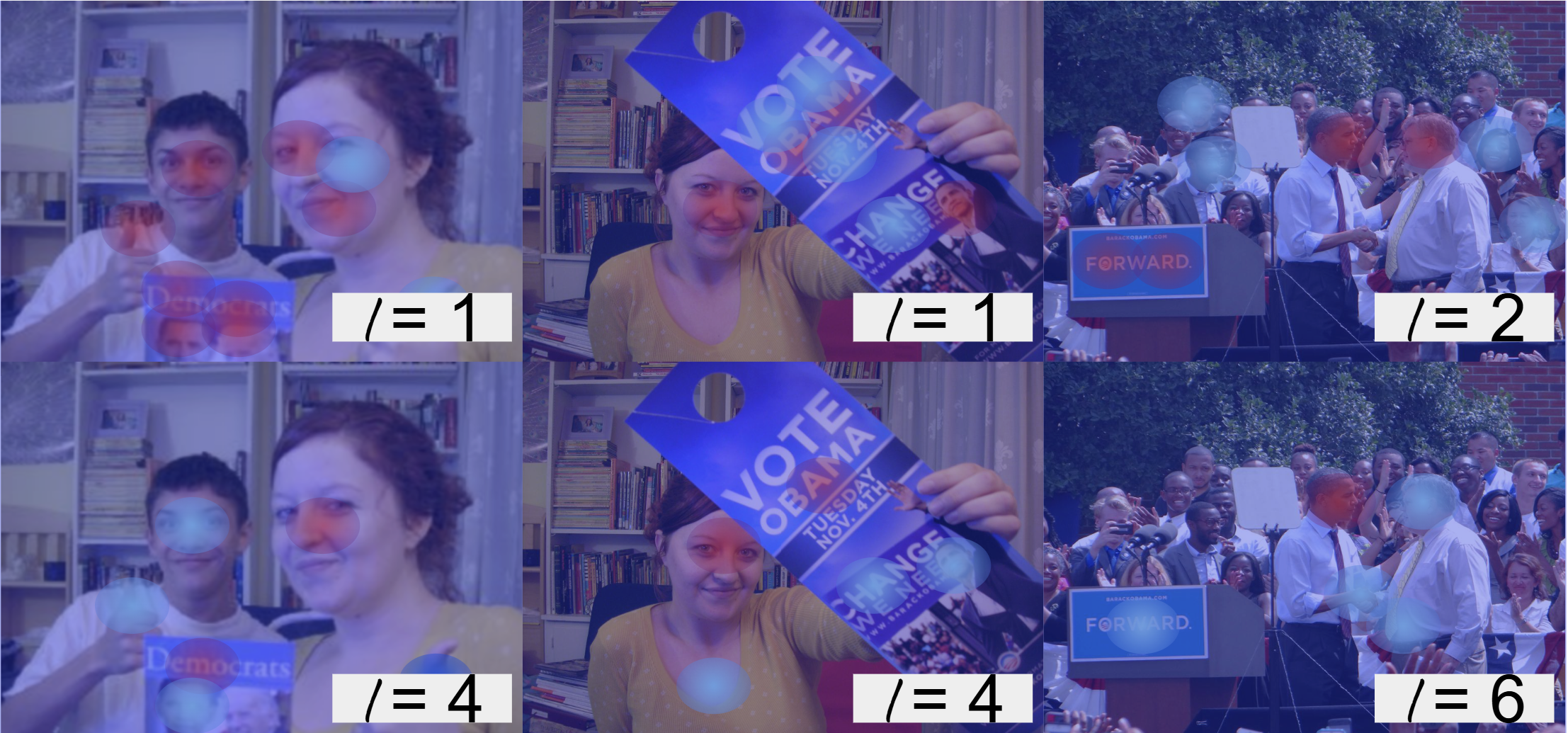}
  \caption{Qualitative example of the inter- and intra-person differences: Each row corresponds to a different participant, where red dots depict the areas that participants attended to the most (i.e. more fixations).
  The user-selected privacy levels $l$ are consistent for each participant but different across participants. For example, the \textit{political opinion} attribute is private ($l$ between 1 and 2 - the private levels) for one user -the first row- and not private for the second ($l$ between 4 and 6 - the safe levels). The gaze fixations are also denser on the private regions of interest (e.g. faces and politicians) for the \textit{private} levels.}
  \label{fig:inter-intra}
\end{figure}

\Cref{tab:spp}\added[id=PR]{ shows the accuracy of the aforementioned models if all (meta)data is passed to the ML model.} Our results demonstrate that gaze data can quantify perceived privacy since results exceed chance levels (0.5 for Task 1 and 0.14 for Tasks 2 and 3). We further analyse this in both person-specific and person-independent settings. 
In each case, models were trained on 80\% of the data and tested on the remaining 20\% with cross-validation.
In the person-specific setting, models are tested on 20\% of individual participants, accounting for personal gaze behaviours \added(with only one sample at a time independent of any user history), whereas in the person-independent approach, a generalizable model is trained across all participants and tested on 20\% of each participant's data. Results in \cref{tab:spp} and \cref{fig:inter-intra} indicate significant variations in inter-user differences (person-independent, leading to lower accuracy), whereas intra-user features exhibit greater consistency (person-specific, leading to higher accuracy). This suggests that each user possesses a distinct and individualised perception of privacy.

\subsubsection*{Fixation-based attention allocation} In general, fixation duration increases on private cues with fewer fixations on less private regions (\added[id=mayar]{$H_0$: \textit{'There is no difference in the distribution of fixation durations across different stimuli regions'}} tested with Kruskal-Wallis H-test and Dunn’s test with Bonferroni correction for multiple comparisons with the different privacy levels, with $p\text{-}value = \num{0.04} < \num{0.05}$ \added[id=mayar]{and Epsilon-squared effect size for Kruskal-Wallis H-test of $0.05$}). This can also be seen in \cref{fig:inter-intra}.
It further suggests that less fine-grained tracking is sufficient (i.e. relying on fixations alone). This was further supported by downsampling the gaze data to \SI{30}{Hz} (\SI{30}{FPS}) to simulate commodity-level standard webcams, as shown in \cref{tab:spp}-downsampling. The results, again, suggest that high-resolution gaze data (\cref{tab:spp}-person-independent) may not be necessary for effective privacy perception prediction, facilitating a more widespread implementation and enhancing accessibility and usability for a broader audience \footnote{As this study represents exploratory research, we employed a high-resolution, constrained experimental setup. This choice was motivated by the initial uncertainty regarding whether a lower-resolution configuration would yield significant results and to what extent fine-grained details in the data would be necessary to capture relevant effects. The high-resolution setup ensured maximal data fidelity, allowing for comprehensive observation of potentially subtle phenomena during the early stages of research.}. 

\subsubsection*{Social influence:} 
\begin{figure}
\centering
  \includegraphics[width=0.45\textwidth]{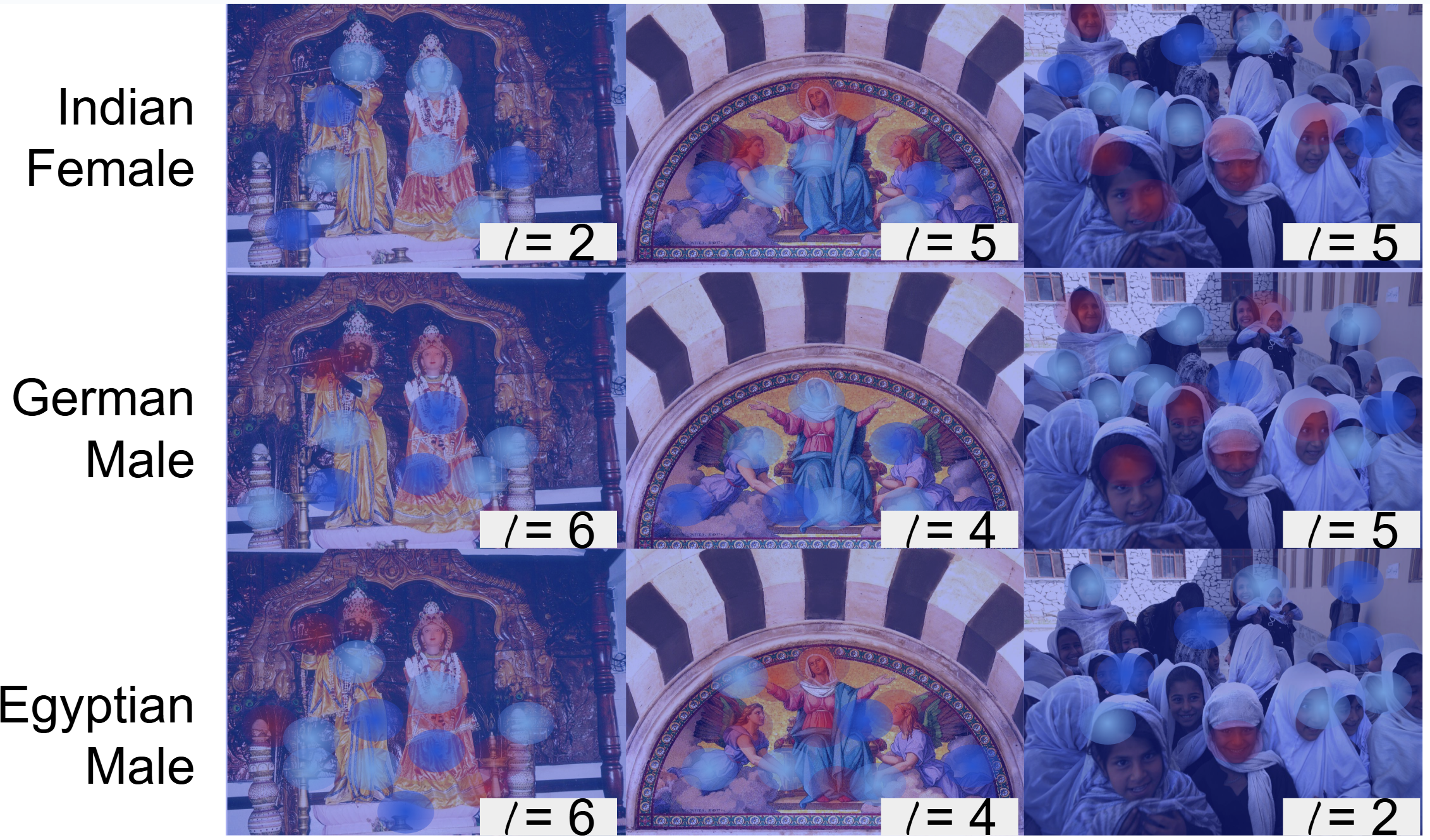}
  \caption{Qualitative example of social influence: By looking at the attribute \textit{religion}, participants assign \textit{private} ratings to stimuli that are more closely associated with or frequently encountered within their social background, probably due to their personal significance (c.f. \cref{sec:related_work}).}
  \label{fig:nationality}
\end{figure}

Users’ privacy perception may be guided by social influence (e.g. demographics), as shown in the \textit{Contextual Privacy Perception} results and \cref{fig:nationality}.
Our results indicate that demographics play a significant role in shaping outcomes (i.e. increasing accuracy). This suggests that demographic factors and social background contribute to identifiable patterns that machine learning models can detect. Consequently, these factors enhance the predictive power of the models, highlighting their relevance in understanding variations in the data.\footnote{It is important to note that our unbalanced demographics—an inherent consequence of the open and uncontrolled call for participation—limit our ability to make broad or generalizable claims about specific demographic groups or extrapolating results to a wider population. As a result, while our analysis may reveal meaningful trends, caution is required when interpreting demographic-specific conclusions using our dataset}

\subsubsection*{Long-term learning and adaptive privacy behaviour} 
\begin{figure}
\centering
  \includegraphics[width=0.6\columnwidth]{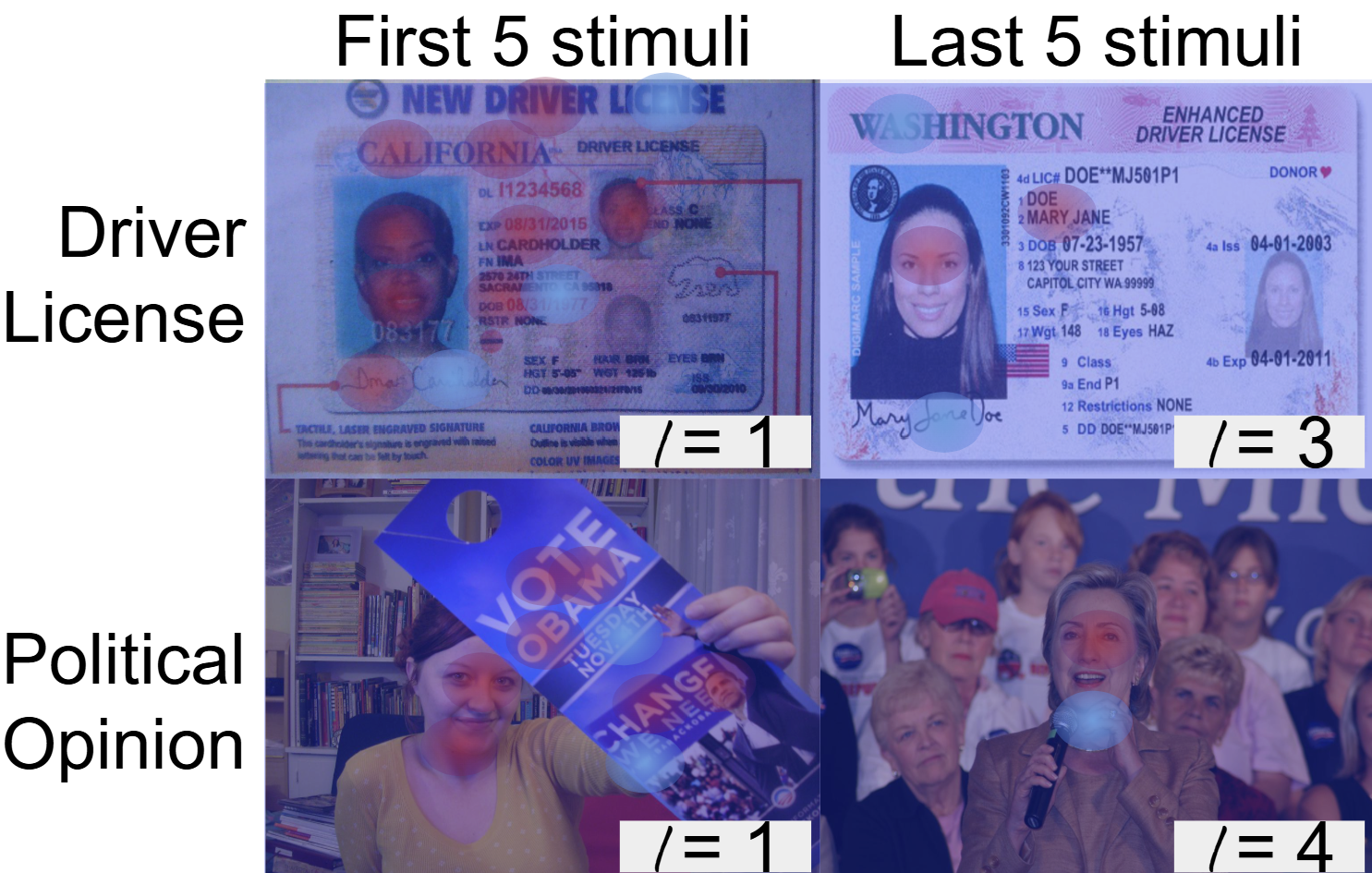}
  \caption{Qualitative example of learning effect: 
  Examples belong to the same participants on images of two different attributes \textit{driver license} and \textit{political opinion}, sampled from the first and last 5 stimuli. When a specific attribute is presented (e.g. in the first 5 stimuli) and repeated multiple times, it increases the participant's familiarity with the attribute through repeated exposure. Hence, the number of fixations, response time, and privacy-sensitivity decrease.}
  \label{fig:learning}
\end{figure}

Repeated exposure to private attributes changes user behaviour over time. As shown in \cref{fig:learning}, while gaze fixations decrease on repeated attributes over time (i.e. desensitisation to privacy risks), the privacy ratings increase (i.e. learning effect and growing privacy awareness), and response time decreases.

\subsubsection*{Visual Privacy}
\begin{figure}
\centering
  \includegraphics[width=0.5\textwidth]{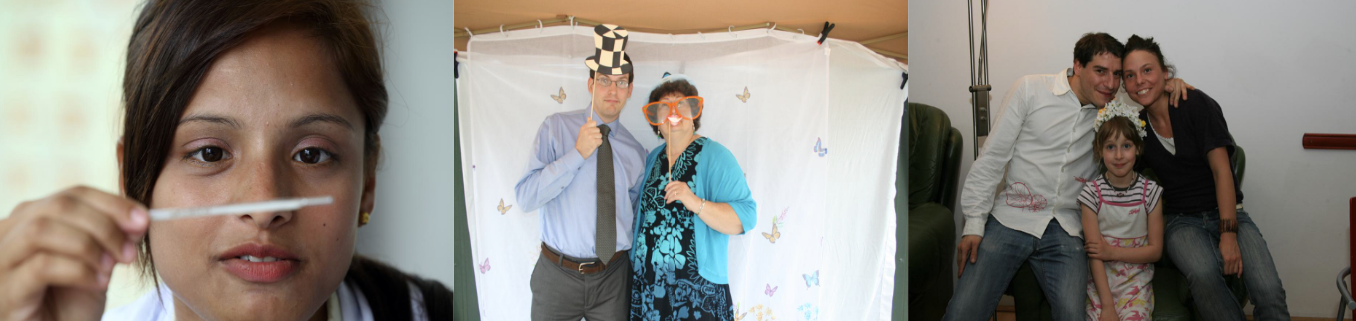}
  \caption{Qualitative example for visual privacy observations: Example stimuli corresponding to the gaze data that were misclassified}
  \label{fig:spp_ablation}
\end{figure}

For private attribute recognition, as shown in \cref{tab:attribute}, we compared our findings with the VISPR \cite{orekondy2017advisor} evaluation \added[id=mayar]{as well as other SOTA models}. While \cite{orekondy2017advisor} only input the stimuli, we replicated their models and metrics on our dataset and inputted the gaze data alone vs the gaze and stimuli. 
%As shown in \cref{tab:attribute},  when used in conjunction with stimuli, gaze data enhances the informational richness available to the model by providing additional features that complement the stimuli. 
%However, even when gaze data is used in isolation, it remains beneficial while reducing the amount of shared data. Note that the models used are relatively basic for the sake of comparison with \cite{orekondy2017advisor}, and more advanced models could potentially enhance the results. 
Nonetheless, we observed that, as shown in \cref{fig:spp_ablation} (left to right): (i) when using gaze as the only input, the models are able to identify objects that occupy a significant proportion of the image space while failing at identifying other subordinate attributes such as age or gender; (ii) when augmenting the stimulus with the gaze data, the models are able to identify more fine-grained details such as wedding rings and tattoos that the stimuli-alone version fails at (as long as the participants paid attention to such details), (iii) in all versions, models fail at identifying the relationship-based attributes (as also reported by \cite{orekondy2017advisor}) since they require some reasoning capabilities that our basic models do not achieve.

\begin{table}
  \caption{Evaluation of the private attribute recognition task as 
  class-based mean average precision (C-MAP) (the average of the per-attribute average of the area under the Precision-Recall curve on all attributes), following \cite{orekondy2017advisor}'s models (CaffeNet, GoogleNet, and ResNet) \added[id=mayar]{and more recent SOTA models (Multimodal fusion model: GazeFormer \cite{mondal2023gazeformer}, Cross-modal transformer: ViLT \cite{kim2021vilt}, and self-supervised contrastive model: CLIP \cite{radford2021learning})}}
  \label{tab:attribute}
  \begin{tabular}{ccccc}
    \toprule
    \textbf{Model} & \textbf{Feature} & \textbf{\makecell{VISPR \cite{orekondy2017advisor} \\ (stimuli)}}   & \textbf{\makecell{Ours \\ (gaze)}}  & \textbf{\makecell{Ours \\ (gaze + stimuli)}}  \\
    \midrule
    SVM & CaffeNet & 41.34  & 29.80 & 58.22  \\
    SVM & GoogleNet & 43.77 & 30.07 & 60.43 \\
    SVM & Resnet-50 & 44.21 & 32.45 & 62.37 \\
    \midrule
    E2E & CaffeNet & 47.56 & 35.98 & 65.85  \\
    E2E & GoogleNet & 47.72 & 34.08 & 67.00 \\
    E2E & Resnet-50 & 56.13 & 35.32 & 69.13\\
    \bottomrule
    SOTA & GazeFormer & 64.13 & 40.82 & 92.73  \\
    SOTA & ViLT & 66.18 & 41.59 & 89.47 \\
    SOTA & CLIP & 61.00 & 35.06 &   78.56\\
  \bottomrule
\end{tabular}
\end{table}

\subsubsection*{Gaze as a supplementary input} We assess model performance with the inclusion of stimulus data and compare it to PrivacEye \cite{steil2019privaceye}. Results in \cref{tab:spp}-gaze+stimuli show that the stimulus features significantly improve the performance.
Nonetheless, models trained solely on gaze data still yield meaningful results. This demonstrates that gaze behaviour alone carries informative signals related to perceived privacy, supporting its utility in scenarios where access to full visual content may be restricted or deliberately excluded for data minimization and privacy-preserving purposes.

\subsubsection*{Bottom-up and top-down attention}
To %\linelabel{q18:1}
\added[id=mayar]{minimise the confounding influence of bottom-up (saliency-driven) gaze, we attempted to control the visual composition of our stimuli by selecting the VISPR images that mostly contain a single dominant attribute. This design choice was intended to reduce the likelihood that gaze behaviour would be driven by reflexive visual attention to salient features, such as bright colours, known to automatically attract attention regardless of the observer’s cognitive or emotional state. Such bottom-up attention is inherently stimulus-driven and unrelated to higher-order constructs like perceived privacy (i.e. top-down attention).
However, this assumption may not hold in all scenarios, especially when stimuli are real-world images. To address this, we check if residual gaze patterns — i.e., the part of gaze behaviour not explained by bottom-up saliency — are more predictive of perceived privacy than raw gaze data. We use a saliency prediction model (DeepGaze \mayar{cite}\cite{deepgaze3}\pr{?}) to generate a saliency map for each stimulus image. This represents the expected gaze distribution if attention were purely bottom-up. Then we compute the residual gaze maps by subtracting the (normalised) saliency map from the original gaze map. We input both the original gaze maps as well as the residual gaze maps to the models, as shown in \cref{tab:spp}. Results are significantly lower for residual gaze maps, which may indicate that privacy-relevant content often coincides with visually salient regions. Additionally, the lack of variance in saliency features reduces the need for correction, leading to differences in performance between the two models, since the stimuli used in the study were relatively homogeneous in terms of saliency, mostly with one dominant object.
}

\subsection{User-based Perceived Privacy (UPP)}\label{sub:upp}
Even for the same stimuli, different users develop different privacy perceptions.
Hence, perceived privacy is subjective and varies substantially among participants. Understanding the user-specific features of privacy allows researchers and policymakers to design more effective, personalised, and user-centric privacy solutions.
We, therefore, propose three UPP-tasks:

\subsubsection*{{(1) Privacy Expertise Prediction}}
\label{task:knowledge}
Prior works \cite{brams2019relationship, liu2009expert, law2004eye} showed that distinctive eye movement behaviours and gaze strategies correlate with domain-specific expertise.
This task aims to develop a model that can distinguish between privacy experts and non-experts based on their gaze patterns during interactions with digital content or privacy-related tasks. The gaze data was collected from participants with varying levels of privacy expertise, potentially revealing how different levels of knowledge influence visual attention and decision-making in privacy-sensitive contexts. We address this task, therefore, by training a classifier to distinguish between the two groups based on the gaze features. 

\subsubsection*{{(2) User Privacy Profiling}}\label{task: profile}
The privacy expertise prediction task can be further extended to capture the gaze behaviour profile, e.g. a summary of the key gaze features that characterise each group \cite{orekondy2017advisor}, such as differences in attention to specific elements or gaze strategies. This is commonly used to adapt privacy preferences according to different cohorts \cite{google_floc_whitepaper}.
Hence, the task aims to cluster the different groups according to the gaze patterns. 

\subsubsection*{{(3) Privacy-aware Gaze Identification}}

Current gaze-based user identification methods often rely on specially designed visual stimuli or induced artificial gaze patterns. Prior works \cite{djeki2024reimagining, abdrabou2024eyeseeidentity} have investigated the feasibility of distinguishing individuals based on their natural gaze behaviour while freely viewing images and suggested that viewing different images, such as a personal vacation photo, elicits distinct emotional responses, which are reflected in gaze patterns and are unique to each individual. 
We extend this idea to privacy-aware gaze behaviour, proposing that individuals exhibit unique gaze patterns not only in response to personal relevance but also when assessing perceived privacy. Privacy perception is inherently subjective, shaped by personal experiences, cultural influences, and cognitive biases. 
By analysing how users visually explore and evaluate images with different privacy implications, we need to demonstrate in this task that privacy perception itself can serve as an implicit user identifier \cite{d2023using}. 

\bpara{UPP applications} %\linelabel{q15:2} 
User-specific insights can, for example, be used for privacy nudges \cite{ioannou2021privacy} (to encourage users to make privacy-conscious decisions according to their expertise or profiles, commonly used in social media platforms to prevent oversharing, e.g. Facebook’s contextual privacy warnings \cite{meta_sensitive_content_2024} and Chrome browser-based security warnings \cite{google_safebrowsing}), privacy labels and transparency notices (to help users understand and control their privacy choices according to their profiles, e.g. Apple’s App Store privacy labels \cite{apple_privacy_labels}), and cohort-based recommendations (for group-based personalization, e.g. Google’s Federated Learning of Cohorts (FLoC) \cite{google_floc_whitepaper} for ad targeting) (C.f. \cref{apx:apps}). 

\bpara{UPP Baselines and Evaluation}
To predict the privacy expertise of users and groups, we ran the same evaluation as in \cref{sub:spp}. The results are shown in \cref{tab:upp}, where results exceed the chance levels (0.5 and 0.01 accuracy for privacy expertise prediction and privacy-aware gaze identification, respectively). We observe the following key results:

\subsubsection*{Cognitive and Perceptual Adaptation} 
The difference in gaze behaviour between experts and non-experts arises due to cognitive and perceptual adaptations that develop with experience and training \cite{brams2019relationship, liu2009expert, law2004eye}.
Our results show that privacy experts exhibit more focused attention on privacy-relevant information and quicker identification of potential privacy risks compared to non-experts.
In other words, privacy experts have rapid gaze shifts, suggesting automatic heuristic-based decision-making, while non-experts have longer dwell time, indicating uncertainty and high cognitive load in privacy assessment (\added[id=mayar]{$H_0$: \textit{'The gaze duration distributions of the two groups (experts and non-experts) are equal'}} tested by a Mann-Whitney U-test with  $p\text{-}value = \num{0.03} < \num{0.05}$ \added[id=mayar]{and effect size $r = 0.3$}), leading to distinguishing features that ML models can identify with high accuracy (c.f. \cref{tab:upp}).

\subsubsection*{Distinct Profiles} 
We further evaluated the participants' profiles per attribute and found out that, following \cite{orekondy2017advisor}, K-means clustering yields the lowest silhouette score with 12 distinct gaze behaviour profiles (as opposed to 30 profiles when clustering images \cite{orekondy2017advisor}).

\subsubsection*{Genuine significance}
Previous studies on gaze behaviour indicate that viewing different images, such as personal photos, elicits distinct emotional responses that are reflected in gaze patterns and are unique to each individual \cite{djeki2024reimagining, abdrabou2024eyeseeidentity}.
In our experimental setup, we tried to simulate this setup by instructing participants to conceptualise the stimuli as their own phone gallery without incorporating actual personal photos.
Unfortunately, the previously observed personalised effect could not be replicated in our study and setup (cf. \cref{tab:upp}).
We believe that this is due to the lack of genuine emotional attachment and/or lack of personal significance and familiarity.

\begin{table}
  \caption{Accuracy of the UPP tasks}
  \label{tab:upp}
  \begin{tabular}{lcc}
    \toprule
      & \textbf{\makecell{Privacy Expertise \\ Prediction}} & \textbf{\makecell{Privacy-aware Gaze \\ Identification}}  \\
    \midrule
    DT & 0.78 &  0.05\\
    SVM & 0.45 & 0.04\\
    LR & 0.46 & 0.02\\
    RF & 0.87  & 0.04\\
    KNN & 0.85 & 0.03\\
  \bottomrule
\end{tabular}
\end{table}

%% file: sections/applications.tex
Our results so far showed how gaze can be used to predict the privacy perception of an individual or a group of users.
There are many potential applications for this approach (we already hinted at some of them).
Here, we concentrate on one important task, and study this in more detail, namely, finding good and tailored epsilon values for DP.

\subsection{Privacy-Utility-Cognition Trade-Off}\label{sec:priv-util}

To use our results from \cref{sec:benchmarks} with a DP-mechanism, we first need to map user-perceived privacy levels $l \in \{1, \ldots, L\}$ with a function $f$ to $\varepsilon$-values for Differential Privacy.\footnote{In our evaluation, we use $L=7$ (cf. \cref{sec:related_work}).} 

Depending on the cognitive aspects behind the user-perceived privacy preference, which we discussed already in \cref{sec:related_work}, we want to connect $l$ with different privacy loss functions $g(l)$, e.g. such that the privacy loss depends linearly on $l$ as in \cite{kohli2018epsilon, lu2023improving}.

To construct a suitable mapping $f$ to DP-privacy levels $\varepsilon$, recall from \cref{sec:preliminaries} that a privacy-budget $\varepsilon$ ensures that an adversary $\mathcal A$ can distinguish two data sets with advantage bounded by $\operatorname{adv}_{\mathcal A}(\varepsilon)\coloneqq \frac{e^{\varepsilon}-1}{e^{\varepsilon}+1}$.
Hence, if we require the privacy loss (in terms of the success of a potential adversary)
to have a certain behaviour $g$, we need $g(l)=\operatorname{adv}_{\mathcal A}(f(l))$. 
Observe that we can compute $f(l)=\log\left(\frac{1+g(l)}{1-g(l)}\right)$ explicitly.
In particular, we find a mapping function $f$ for every positive loss function $g$ with values smaller than $1$.

Since users do not select privacy levels purely based on objective risks or technical parameters, but rather through subjective interpretations shaped by cognitive biases (cf. \cref{apx:theories}), we now want to discuss the typical choices of $g$ in cognitive science and how they affect the accuracy of a model obfuscated with $\varepsilon$-DP noise if $\varepsilon=f(l)$.

Note that for real-world applications, we can usually restrict the privacy budget to $\varepsilon_{min}\leq \varepsilon\leq \varepsilon_{max}$ where the minimum $\varepsilon_{min}\geq 0$ and maximum $\varepsilon_{max}<\infty$ are application-specific \cite{nasr2021adversary}. 
We therefore, use functions $g$ depending on $\varepsilon_{min}, \varepsilon_{max}$.

It is important to note that users employ diverse rationales and cognitive strategies when forming their privacy perceptions, and these strategies are shaped by multiple interacting factors, such as the stimuli, background knowledge, and demographics (cf. \cref{sec:benchmarks}).
As a result, there is no universally correct mapping function, as these differences reflect valid subjective interpretations rooted in cognitive and contextual variability. Rather than assuming a single ground truth, such mappings should be empirically learned from user behaviour and validated through data-driven analysis.
Hence, later in this section, we illustrate this point by providing a practical example based on our dataset.

\subsubsection{Linear Mapping} 
A straightforward way to interpret perceived privacy is a function $g$ linear in $l$ \cite{kohli2018epsilon, lu2023improving}.
Namely, we choose for $l\in\{1,\ldots, L\}$: 
\begin{align*}
g(l) = \operatorname{adv}_{\mathcal A}(\varepsilon_{min}) + \frac{l-1}{L-1} (\operatorname{adv}_{\mathcal A}(\varepsilon_{max}) - \operatorname{adv}_{\mathcal A}(\varepsilon_{min}))
\end{align*}

This mapping assumes equal intervals between privacy levels. Linear mappings are widely used in statistics, cognitive sciences, and machine learning for their mathematical simplicity, speed, and interoperability, and serve as good approximation functions. Nonetheless, they may potentially lead to inaccurate interpretations \cite{burkner2019ordinal}. 
This is because, although privacy levels are numerically treated as equidistant, the psychological or cognitive perception of the distance between those points may not be equal \cite{waldman2020cognitive, kezer2022getting}. For example, the difference between 'very safe' and 'safe' may be much smaller in the user’s mind than the difference between 'neutral' and 'moderately private'. Similarly, the midpoint ('neutral') may be seen not as a numerical centre but rather as a safe choice. Hence, we no longer have a linear relation.

\begin{figure*}
\centering
  \includegraphics[width=1\textwidth]{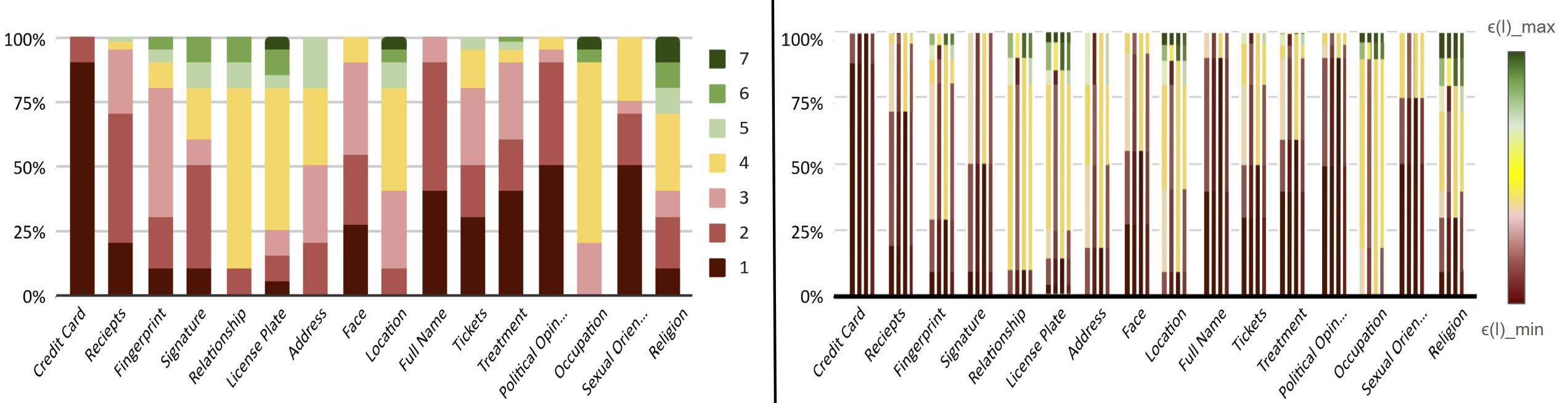}
  \caption{The left figure shows the distribution of the perceived privacy levels $l \in$ \{1 (very private),...,7 (very safe)\} selected by the participants per attribute. The right figure shows an example of mapping the perceived privacy level $l$ to the corresponding $\varepsilon(l)=f(l)$ using the different mapping functions for $\varepsilon_{min} = 0.1, \varepsilon_{max} = 5$ and $k= 1.5$. The results are applied to the $l$ distribution where, for each attribute, the mapping functions (linear, exponential, sequential, and sigmoid) are depicted from left to right.}
  \label{fig:dist}
\end{figure*}

\subsubsection{Exponential Mapping} Alternatively, an exponential function could be used when users interpret privacy levels as a continuum, with thresholds separating the different privacy levels. In other words, in such cases, small variations in the privacy parameter may be perceived as insignificant at higher levels (e.g. 'very safe' and 'moderately safe'), while similar changes near specific lower thresholds elicit disproportionately larger responses (e.g.'neutral' and 'moderately private'). These thresholds may vary due to the user's cognitive processes, where users with less familiarity with privacy concepts may rely on intuitive or categorical decision-making or cases where specific features of the images (e.g., presence of identifiable faces or objects) might consistently evoke higher or lower ratings. 
An exponential mapping addresses this pattern and emphasises stronger privacy guarantees at lower ratings and captures a cognitive tendency where users may undervalue small privacy differences in higher ratings, improving utility. We get:
\begin{align*}
    g(l) = \operatorname{adv}_{\mathcal A}(\varepsilon_{min})^{(L-l)/(L-1)}\operatorname{adv}_{\mathcal A}(\varepsilon_{max})^{(l-1)/(L-1)}
\end{align*}
\subsubsection{Sequential Mapping} Another alternative is a sequential mapping, where the choice of the privacy level is reached in steps. This models the probability of the selected privacy level being in a particular category, given that it has surpassed the previous category. 
In some cases, decision-making unfolds in steps \cite{nissenbaum2004privacy, kokolakis2017privacy, aivazpour2017unpacking, gharib2024towards}, such as deciding whether something is private and then determining the degree of privacy, e.g. "Is the stimulus sensitive/ does it violate a norm/should the information be shared? If yes, how private are the stimuli?/ How severe is the violation?/ How much information should be disclosed?". In these cases, automatic and heuristic-driven decisions (e.g., 'private or not') may precede more deliberative evaluations of privacy levels \cite{aivazpour2017unpacking, gharib2024towards}. In other cases, the stimuli's complexity may drive such behaviour, e.g. images with progressively more sensitive content may encourage stepwise evaluation (e.g., abstract shapes $\rightarrow$ objects $\rightarrow$ faces).
Therefore, prior works \cite{northoff2004cortical, rolls2004functions, phelps2005contributions} typically model this behaviour using step functions, incorporating heuristic-based decision-making as the primary mechanism to mitigate ambiguity aversion and prevent cognitive overload during deliberative evaluations.

Hence, a piecewise mapping (i.e. a step function) can be used to accommodate the different perceived privacy ranges to handle that different tiers of privacy ratings have different baseline $\varepsilon$ values and slopes and allow more granularity within $n$ ranges. 
Especially when certain ranges require higher privacy guarantees (e.g. handling extreme outliers).
E.g. a look-up table where $n = 3$ (e.g. 'private', 'neutral', 'safe'):

\begin{align*}
    g(l) = \begin{cases}
    \operatorname{adv}_{\mathcal A}(\varepsilon_{min}), & \text{if } l = 1 \text{ or } 2\\
    \frac{1}{2}(\operatorname{adv}_{\mathcal A}(\varepsilon_{max}) + \operatorname{adv}_{\mathcal A}(\varepsilon_{min})), & \text{if }2 < l < L-1\\
    \operatorname{adv}_{\mathcal A}(\varepsilon_{max}), & \text{if } l = L \text{ or } L-1
\end{cases}
\end{align*}

\subsubsection{Sigmoid-based Mapping} 
Furthermore, adjacent-category models \cite{burkner2019ordinal} are common in statistics and item response theory. They are usually used when thinking of a natural cognitive process is not possible and decisions may involve iterative, contextual, uncertain, or unstructured processes.
In these cases, users can make fragmented decisions or skip steps due to limited information. In other cases, similar stimuli with only slight changes in features (e.g., cropped vs. full image) require users to make nuanced decisions. 
Commonly in machine learning and statistics, sigmoid (logistic) functions are used to describe the probability of selecting one option over another under uncertainty. It naturally models probabilistic decisions that become more deterministic as the evidence or difference between choices increases.

This behaviour aligns well with empirical findings in psychology, economics, and cognitive science, where decision-making often follows a logistic-like pattern. That is, in stochastic cases, for highly private images where the perceived privacy is greater (i.e. lower ratings), the sigmoid function outputs a lower $\varepsilon$ value, accentuating privacy protection. Conversely, for images considered safe, the function assigns a higher $\varepsilon$, allowing less noise and thus better utility of the data.

Hence, a sigmoid mapping function can be expressed as the smooth continuation of
\begin{align*} 
    g(l) =  
    \operatorname{adv}_{\mathcal A}(\varepsilon_{max}) +\frac{\operatorname{adv}_{\mathcal A}(\varepsilon_{min}) - \operatorname{adv}_{\mathcal A}(\varepsilon_{max})}{1+\left(\frac{L-1}{l-1}-1\right)^{-k}}
\end{align*}
where $k$ controls the steepness of the curve (higher values make the transition sharper, e.g. $k=1.5$ represents moderate steepness). 
As shown in \cref{fig:dist}, such smooth transitions ensure that $\varepsilon$ values transition gradually instead of changing too sharply with diminishing effects at the extremes, allowing to smoothly handle uncertainty or borderline cases without abrupt changes in behaviour.

\subsection{Empirical Evaluation}

%\rk{This section should start with something like: We now evaluate our results from Section 5 in combination with  mappings to epsilon values proposed in Section 6.1 We evalute this as follows ....}

%\mayar{TBD: Reviewer D: Conduct a small-scale qualitative or quantitative study in which users are shown epsilon-calibrated outputs (e.g., noisy queries, masked samples, aggregated stats), and asked whether the level of protection feels “appropriate” given the content.}

We now show how our results from \cref{sec:benchmarks} could be used in combination with mappings to epsilon values proposed in \cref{sec:priv-util}. To illustrate the applicability of our approach, we adopt a representative example from the \textit{privacy level perception} task. This example serves to show how our method can be operationalised in practice. Importantly, the proposed framework is not limited to this specific task; it can be extended to other tasks by incorporating stimuli-or user-based information, thereby enhancing its generalizability across various privacy-sensitive applications.

We apply the state-of-the-art personalised differential privacy (PDP) mechanisms for privacy budgets $\varepsilon$ determined with our mapping functions from \cref{sec:priv-util}.
%Recall that \cite{jorgensen2015conservative} is originally used with randomly chosen $\varepsilon$.
%\linelabel{q19:2}
\added[id=mayar]{We employ the existing PDP methods without modification to their core mechanisms. The only alteration involves replacing the randomly generated privacy-level inputs ($\varepsilon$) commonly used in prior work with user-specified privacy preferences inferred from gaze. This allows for a more realistic and user-informed evaluation of PDP behaviour while preserving the original algorithmic structure.}
For our comparison, we use different common tasks from data analysis and machine learning tasks, which we want to describe briefly next. 
%This is based on prior works \cite{sattar2020deep, sattar2015prediction} that were able to infer some general properties (e.g. colours) of a search target. They further proved that the search task is guided by the user's mental model and showed that mental models substantially differ among participants. In addition, when individuals perceive that their privacy is at risk, their intentions to engage, share information, or use certain services are often negatively impacted \cite{malhotra2004internet, dinev2006extended, krasnova2010online}.
%We, therefore, hypothesise that privacy perception significantly influences individuals' intentions (i.e. search targets).
%\rk{For me it does not become clear what the evaluation is about and how our previous results are used. And what has the above mentioned, mental model to do with all the rest? }
%\pr{I did not see the connection to the benchmarks either, so removed this part.}\mayar{clarified above.}

\paragraph{Benchmarks} 
Our evaluation uses the same benchmarks as Jorgensen et al. \cite{jorgensen2015conservative} to evaluate PDP mechanisms for search analytics, e.g. count query for the number of documents that participants searched for, median and minimum queries for attention allocation (number of fixations) per stimulus. 
Once a $\varepsilon$-value is determined, either randomly by \cite{jorgensen2015conservative} or by applying our method on the user's gaze data, %In this setup, each participant specifies the \rk{Isn't the point of our approach to figure our individual epsilon values?}\pr{I also did not understand what was happening here -- I made an educated guess. Mayar, please change this if it is wrong.}
%$\varepsilon$ value and sends their data to a trusted data analyst. 
a trusted data analyst  receives the data and the $\varepsilon$ values, then adds suitable noise to query results and publishes the aggregate statistics.
In addition to these classical data analysis benchmarks, we also evaluate PDP mechanisms for machine learning tasks, namely, search intention prediction, using the default models and parameters of \cite{jorgensen2015conservative, boenisch2023have, boenisch2022individualized} (cf. \cref{sec:related_work}).

\begin{table*}[t]
  \caption{Evaluation of the PDP benchmarks. The table shows results for data analysis and machine learning tasks where (i) plain is the non-private computation and, hence, the best utility, (ii) static is the worst-case privacy loss and the most commonly used in standard DP protocols, (iii) random is a random distribution of $\epsilon$ values that are commonly used in PDP protocols, and (iv) our four proposed mapping functions.}
  \label{tab:pdp}
  \begin{tabular}{llccccccc}
    \toprule
    & & \textbf{Plain} & \textbf{Static} & \textbf{Random} &  \textbf{Linear}  & \textbf{Exponential} & \textbf{Sequential} & \textbf{Sigmoid} \\
    \midrule
    \textbf{Analysis} & Count \cite{jorgensen2015conservative} & 100 & 76 & 84 & 87 & 90 & 86 & 87 \\
     & Median \cite{jorgensen2015conservative} & 16 & 12 & 19 & 14 & 18 & 19 & 19 \\
     & Min \cite{jorgensen2015conservative} & 5 & 8 & 7 & 7 & 4 & 7 & 6 \\
    \midrule
     \textbf{Learning} & Linear regression \cite{jorgensen2015conservative}
     & 0.57 & 0.31 & 0.43 & 0.46 & 0.50 & 0.49 & 0.47\\
     & Weighting \cite{boenisch2022individualized}
     & 0.68 & 0.38 & 0.52 & 0.53 & 0.57 & 0.56 & 0.55\\
     & Weighting \cite{boenisch2023have}
     & 0.49 & 0.32 & 0.43 & 0.44 &  0.46 & 0.45 & 0.42 \\
     & Sampling \cite{boenisch2022individualized}
     & 0.68 & 0.38 & 0.55 & 0.58 & 0.60 & 0.61 & 0.63\\
     & Sampling \cite{boenisch2023have}
     & 0.49 & 0.32 & 0.43 & 0.43 & 0.44 & 0.44 & 0.46\\
  \bottomrule
\end{tabular}
\end{table*}

\paragraph{Results} As shown in \cref{tab:pdp}, PDP approaches significantly improve over the static DP approaches (i.e.
%\linelabel{q19:1}
\added[id=mayar]{the worst-case privacy guarantee by uniformly applying the maximum privacy level required across all users, without adapting to individual preferences or contexts to address the most stringent privacy demands}, the common approach in DP) and are significantly closer to the plain approaches (i.e. without noise addition). In addition, our proposed mapping functions yield better utility than the random benchmark (i.e. the random $\varepsilon$ values generated by the existing benchmarks). 
\newline
To further analyse the effect of the different mapping functions, \cref{fig:dist} shows the resulting $\epsilon$ values for different image attributes. 
The amount of added noise (according to the $\varepsilon$ values) differs according to the mapping function (i.e. more noise is introduced to the private levels represented in red, whereas less noise is applied to those depicted in green), hence improving utility compared to static worst-case approaches.
%\pr{(cf. \cref{apx:pdp_res} for further qualitative results)--currently not defined}.\mayar{I removed the qualitative results, I believe the current ones are sufficient and less confusing.} 
Our results indicate that: 
\begin{itemize}[leftmargin = *]
    \item \textbf{The linear mapping} is particularly appealing in scenarios where exact precision is not critical, and approximate representations are sufficient for the intended application, e.g. when ratings are equally distributed, unlike our dataset, where the number of classes is skewed toward the private ratings. In former cases, linear functions offer a simpler and more computationally efficient mapping. However, the trade-off lies in the potential loss of granularity or accuracy.
    
    \item  \textbf{The exponential mapping} is mostly suitable for applications where strong privacy is paramount and small changes in high ratings imply steep privacy needs. This is true for clearly sensitive attributes, i.e. attributes with an $l$ range $\leq$ 2, such as credit cards. This could also be seen in the higher performance for weighting algorithms in \cref{tab:pdp} where less private data contributes more (given the skewed data distribution as shown in \cref{fig:dist}) with less noise to the final learning outcome.

    \item \textbf{The sequential mapping} demonstrates superior performance for attributes with deliberative privacy sensitivity ($2 < range(l) < 5$), such as political opinions, where participants' gaze shifts between the privacy levels before selection, suggesting an initial classification of the attribute as private, followed by a secondary assessment of its degree of privacy. 

    \item \textbf{The sigmoid mapping} is most effective for attributes where participants either make arbitrary choices (can also be seen in minimal rating time) or are uncertain (can be seen in extended rating time), resulting in attributes with $range(l) > 5$, e.g., license plates that some participants did not pay attention to (i.e., no fixations) or fingerprints with long fixations. This could also be seen in the better performance of the sigmoid mapping for sampling algorithms in \cref{tab:pdp} where neutral samples are sampled more, balancing the data distribution and the added noise, hence improving utility.
\end{itemize}

\paragraph{User expectation alignment}
To %\linelabel{q20:1}
\added[id=mayar]{assess whether the predictions and protections provided by \methodName~align with user expectations, we conducted a follow-up user validation study with a subset of $N=20$ participants. While our quantitative evaluations above demonstrate the technical validity of \methodName~(e.g., accuracy), the user study %herefore, it is important to 
examines the users' subjective acceptance of \methodName.}

\begin{figure}
    \centering
    \includegraphics[width=1\linewidth]{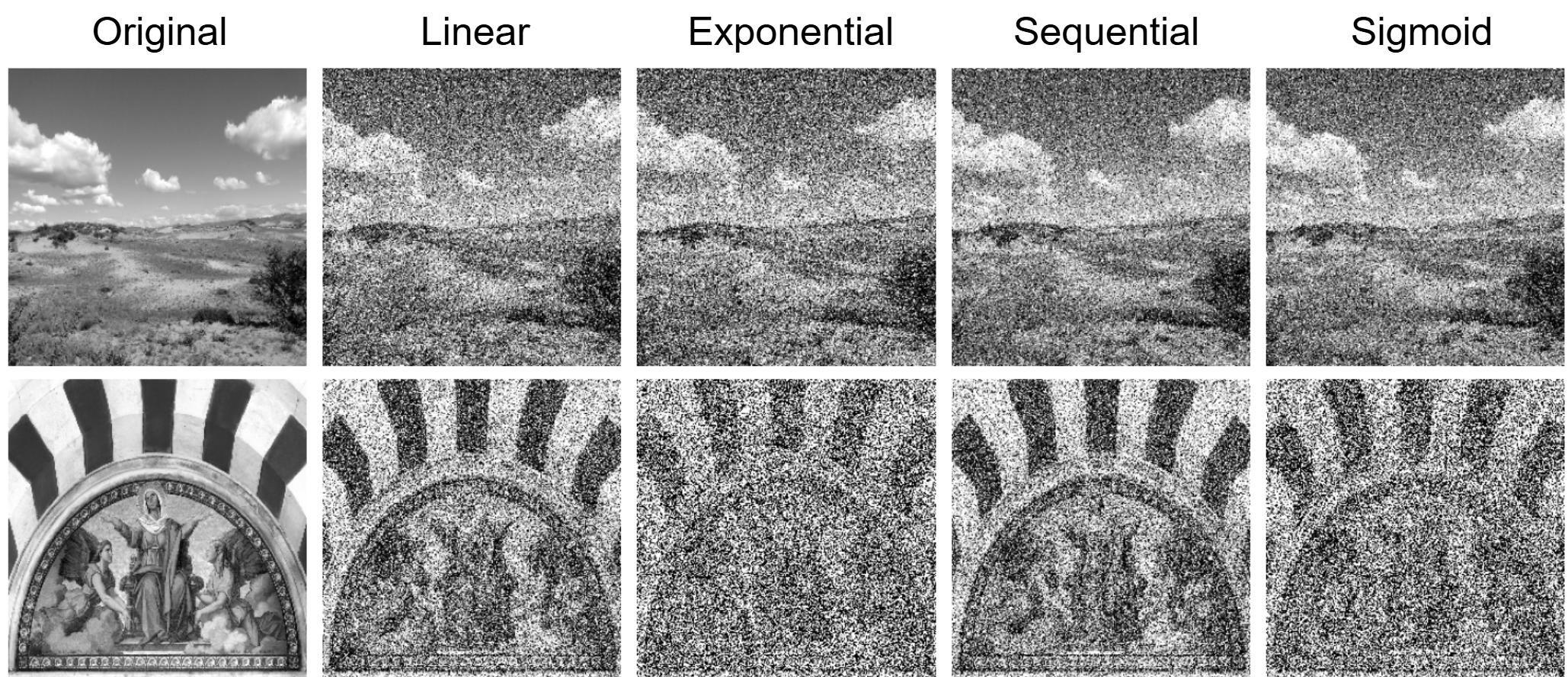}
    \caption{Random samples of the reconstructed stimuli (first row $l= 7$, second row $l=1$.}
    \label{fig:reconstruct}
\end{figure}

%We sent a call back to a subset of our dataset participants ($N = 20$).\\
\added[id=mayar]{For the validation study, each participant was shown the same set of images they had rated previously during the initial dataset collection (the search task), along with their user-specified privacy levels. For each image, we generated a reconstructed version using the standard reconstruction attack pipeline from MLDoctor \cite{MLdoctor} to attack the PDP learning models, simulating an adversarial attempt to recover visual data from differentially private representations. These reconstructions, shown in \cref{fig:reconstruct}, reflect what an external observer might infer about the original image when \methodName's predicted privacy levels are used in PDP.}

\added[id=mayar]{Participants were asked three questions measuring (i) utility, (ii) privacy, and (iii) cognition: (i) if the model's search target prediction was correct. Responses were binary ('yes' or 'no').
(ii) whether the reconstructed image matched their privacy expectations, given the privacy level they previously selected. Responses were collected on a 5-point Likert scale, ranging from 'Not at all' to 'Perfectly aligned', and (iii) if they prefer a certain mapping function, given this privacy-utility tradeoff. Responses were again collected on a 5-point Likert scale per mapping function.
We repeated this procedure for each stimulus, leading to each participant being shown $250$ samples (the same $50$ search stimuli that were previously shown during data collection times our $5$ benchmark models).}

\added[id=mayar]{Results show that the subsample shown to participants was representative, and utility evaluation was close to the numbers reported in \cref{tab:pdp} with 0.5 (i.e. 50\%), 0.6, 0.5, 0.7, and 0.5 correct predictions on average for the learning models in the same order.
In addition, with an average rating of $3.8/5$, the participants' privacy expectations were met. Finally, overall, participants preferred the sequential mapping in 68\% of the cases. They preferred the exponential mapping in 85\% of the private images (rated with $l \geq 6$) and the sequential mapping in 52\% of the images rated as 'neutral', which supports our quantitative results. However, the sigmoid mapping was only selected in 7\% of the cases, showing no clear pattern.}

Therefore, given the absence of a universal solution, we recommend selecting the appropriate mapping function based on the specific application and the characteristics of its stimuli.

%% file: sections/discussion.tex
The introduction of Gaze3P marks a pioneering step in privacy perception, advancing beyond traditional questionnaires by leveraging implicit, dynamic signals derived from gaze behaviour, underscoring that users' visual attention can implicitly reveal privacy sensitivities—offering a real-time, non-intrusive window into subjective privacy evaluation.
More concretely, our dataset showed that eye gaze reliably reflects user perceptions across privacy-related tasks \added[id=mayar]{\textbf{(RQ1)}}. Our ML approach further revealed user- and stimulus-based insights that could be used across a variety of applications \added[id=mayar]{\textbf{(RQ2)}}:
We demonstrated the predictive power of gaze for privacy perception, the behavioural and contextual influences of gaze (e.g. fixation patterns, visual attention, demographic context, and expertise) in shaping privacy perceptions, the potential of gaze-based inference and enhancement of privacy-related attribute recognition and user profiling.

By modelling perceived privacy alongside formal guarantees, we improved data utility while (cognitively) aligning with individual privacy expectations \added[id=mayar]{\textbf{(RQ3)}}. Although a single universal mapping between perception and privacy budgets is unlikely to exist due to inter-individual variability, we propose a set of adaptable mapping functions that can be selected contextually based on application needs and behavioural insights.

\subsection{Limitations}

We \added[id=mayar]{acknowledge an age-related bias in our data, as the participant pool was not age-controlled. Despite a publicly announced call, recruitment likely skewed towards university students and staff due to convenience and accessibility. We also collected supplementary standard demographics, including education level, occupation, and field of study. However, our analysis revealed limited variability across these dimensions, as the majority of participants were students enrolled in STEM (Science, Technology, Engineering, Mathematics) disciplines, with only a few exceptions (4 participants). 
Consequently, while the current dataset may not support strong generalization across diverse demographic strata, this metadata, as well as our data collection software, is included in \cite{code}.
We hope that this allows others to extend analyses or augmentations in future work, particularly with a more demographically diverse sample.}

\added[id=mayar]{Moreover, participants were explicitly instructed to treat the images as if they were their own (e.g., from a personal phone gallery), in line with Sattar et al.  \cite{sattar2020deep}. They were also informed that the study focused on privacy, in accordance with the ethical guidelines and prior research practices \cite{orekondy2017advisor, steil2019privaceye}. While we acknowledge the limitations of this setup and the possible biases it may induce, such constraints are typical in early-stage studies within emerging research domains. Nonetheless, this work serves as a foundational step, enabling future research to adopt more comprehensive methodologies and engage more diverse participant groups to improve generalizability and depth of analysis.}

Similarly, another limitation of the present study is the introduction of contextual bias through priming participants on the topic of privacy. This framing was necessary to align participants with the experimental objectives; however, it likely influenced cognitive processing and visual attention during the tasks. As a result, our predictions are not solely inferred from gaze behaviour in a neutral context but rather from gaze patterns shaped by a privacy-salient environment. This contextualization restricts the generalizability of the findings, as gaze allocation strategies may differ when privacy is not explicitly emphasised. Consequently, the reported predictive performance should be interpreted as applying to privacy-aware scenarios, not to all interaction contexts. Future research should address this by employing between-subject designs with primed and non-primed conditions or by modelling contextual factors explicitly, to disentangle intrinsic gaze-based indicators of privacy perception from those induced by experimental framing.

\subsection{Future Work}
\added[id=mayar]{In this paper, we showed that perceived privacy preferences can be inferred from gaze behaviour alone.
While not the main focus of the present analysis, we also observed that contextual data, such as demographics, have the potential to enrich predictions. 
Future work can build on this foundation by incorporating such auxiliary information to enhance both accuracy and personalisation.
Similarly, additional metadata on individual user traits, such as trust propensity, risk perception, and prior exposure to privacy threats, can be collected to enable a deeper understanding of how such latent factors implicitly influence privacy-related behaviours \cite{guerra2023empirical, prange2024not, elbitar2021explanation}.}

Additionally, our proposed dataset includes rich information that we encourage the community to develop upon. 
This can include (i) additional tasks, such as multi-attribute and multi-user correlation analyses (i.e. each group of stimuli was shown to various participants, with each stimulus containing several attributes), 
(ii) tailored models to capture more patterns and enhance our baseline performance, (iii) other DP and privacy-preserving protocols (i.e. adding noise to different privacy units, the fundamental entity whose privacy is protected, user-, label-, feature- or pixel-level DP) that would benefit from the user (cognitive) privacy perceptions without explicit interaction, and \added{(iv) fine-grained eye-tracking analyses to gain a deeper understanding of gaze behaviour in privacy-sensitive contexts. While the present study demonstrates that learned gaze features can serve as effective predictors of perceived privacy, more granular analyses - e.g. gaze entropy, micro-saccade dynamics, and scanpath structure - may reveal subtle cognitive and affective processes underlying privacy perception and could help disentangle the interplay between bottom-up and top-down visual attention in private settings, improving the interpretability of gaze-based privacy models.} 

%% file: sections/conclusion.tex
Given its inherently subjective nature, which varies substantially across individuals, we presented the first large-scale dataset for studying user-perceived privacy.
The dataset encompasses a diverse range of participants, demographic profiles, and visual stimuli.
Using this novel dataset, we demonstrated that eye gaze can serve as a rich source of information on user-perceived privacy across multiple privacy-related tasks.
Gaze behaviour, by providing implicit and dynamic feedback, offers a powerful and promising avenue for enhancing user interaction and overall system usability.
Moreover, by modelling users’ perceived privacy and applying our findings to PDP protocols -- complementing the underlying mathematical and technical privacy guarantees -- we were able to improve data utility while better aligning with users' expectations of privacy. 
As such, our work bridges the gap between technical and usable privacy by aligning theoretical privacy models with user perceptions.
%, \methodName~bridges the gap between technical and usable privacy.

%% file: sections/appendix.tex
\section{Eye Tracking Data}\label{apx:eye}
Here, we present a detailed breakdown of the process of data collection with an eye tracker (EyeLink 1000), starting from participant recruitment to creating a public dataset:

\subsection{Participant Recruitment}
\begin{figure}
\centering
  \includegraphics[width=0.5\textwidth]{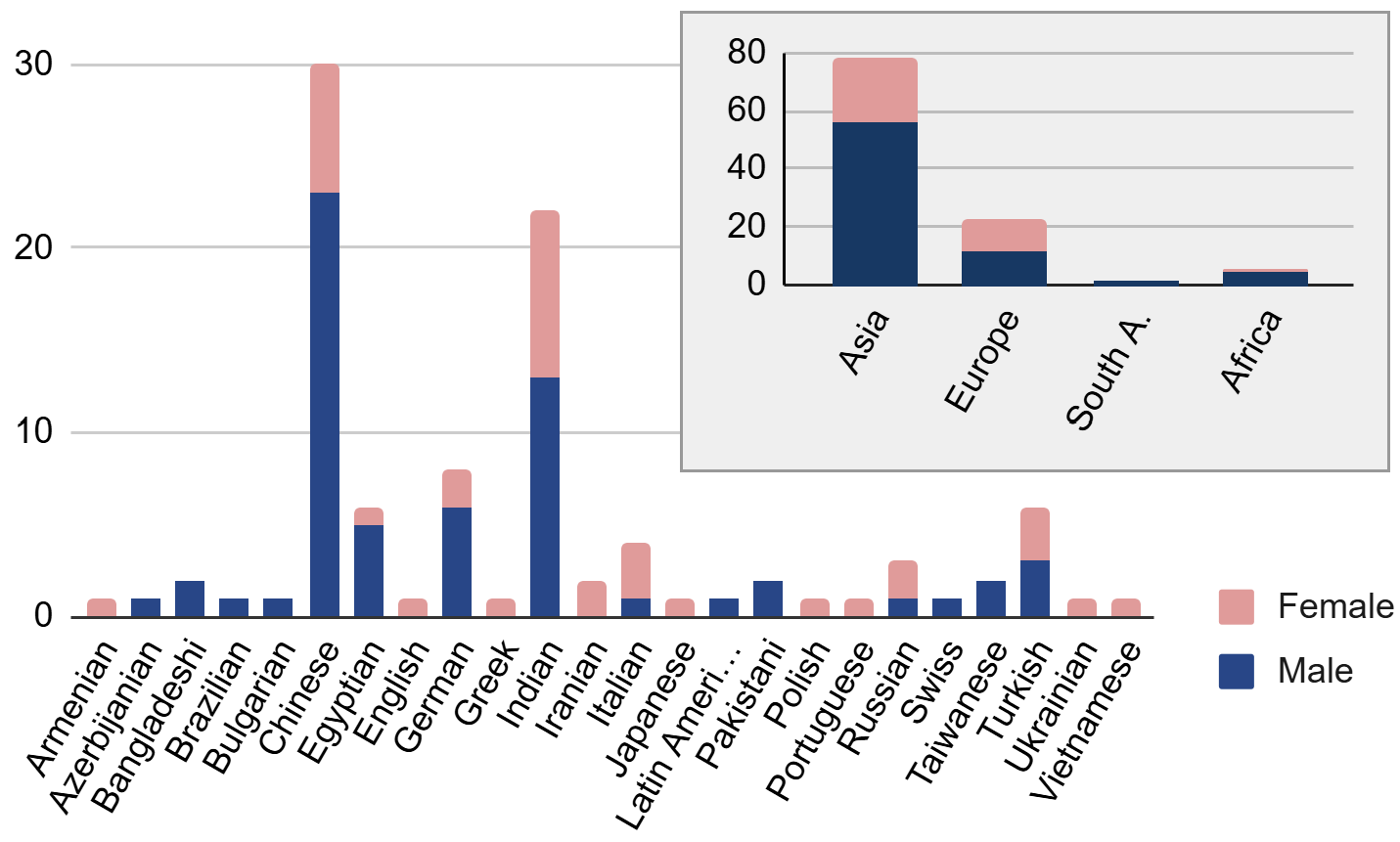}
  \caption{Demographics' distribution of the number of male and female participants per nationality and continent}
  %\Description{...}
  \label{fig:demographics}
\end{figure}

As shown in Figure \ref{fig:demographics}, we aimed to recruit a diverse set of participants with different demographics, e.g. age, gender, nationalities, and AI/Security expertise. 
The call for participation was sent out on different channels, e.g. online platforms, university participant pools, and social media, with clear information about the study, duration, and incentives (e.g., monetary compensation or credits).
Informed consent was obtained following ethical guidelines of the author's institution while ensuring that participants can withdraw at any time.
We excluded participants with specific conditions like eye disorders that may affect tracking accuracy (e.g., nystagmus, extremely poor vision).

\subsection{Eye Tracker Configuration}
\begin{figure}[H]
\centering  \includegraphics[width=0.5\textwidth]{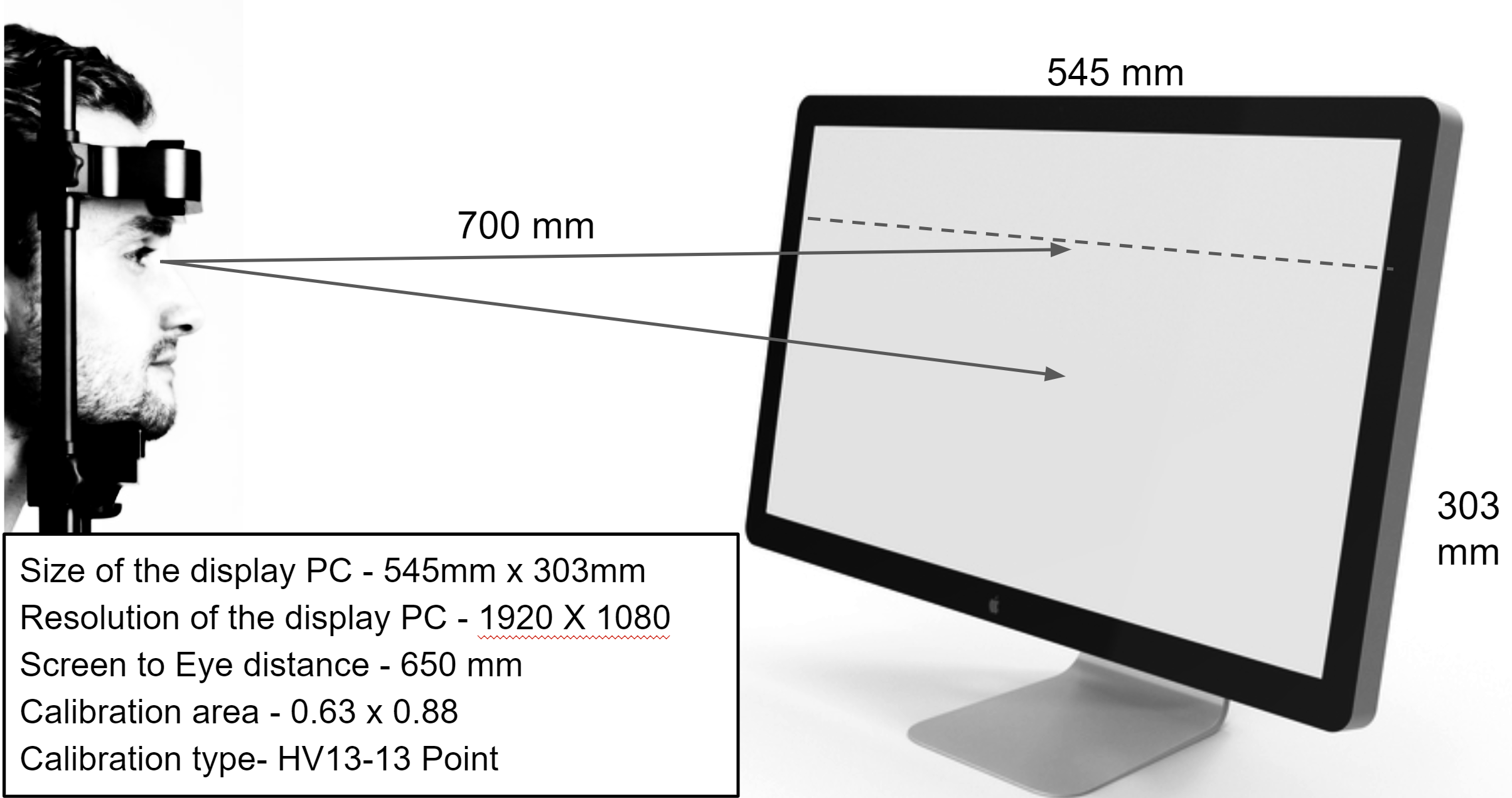}
  \caption{The eye-tracker setup}
  \label{fig:eye_tracker}
\end{figure}

We used the EyeLink-1000 eye tracker, the current state-of-the-art in terms of precision and accuracy for video-based eye tracking. We created an in-lab setup as shown in Figure \ref{fig:eye_tracker}.

\subsection{Experiment Design}\label{app:experimental-design}
\begin{figure}[H]
\centering  \includegraphics[width=0.35\textwidth]{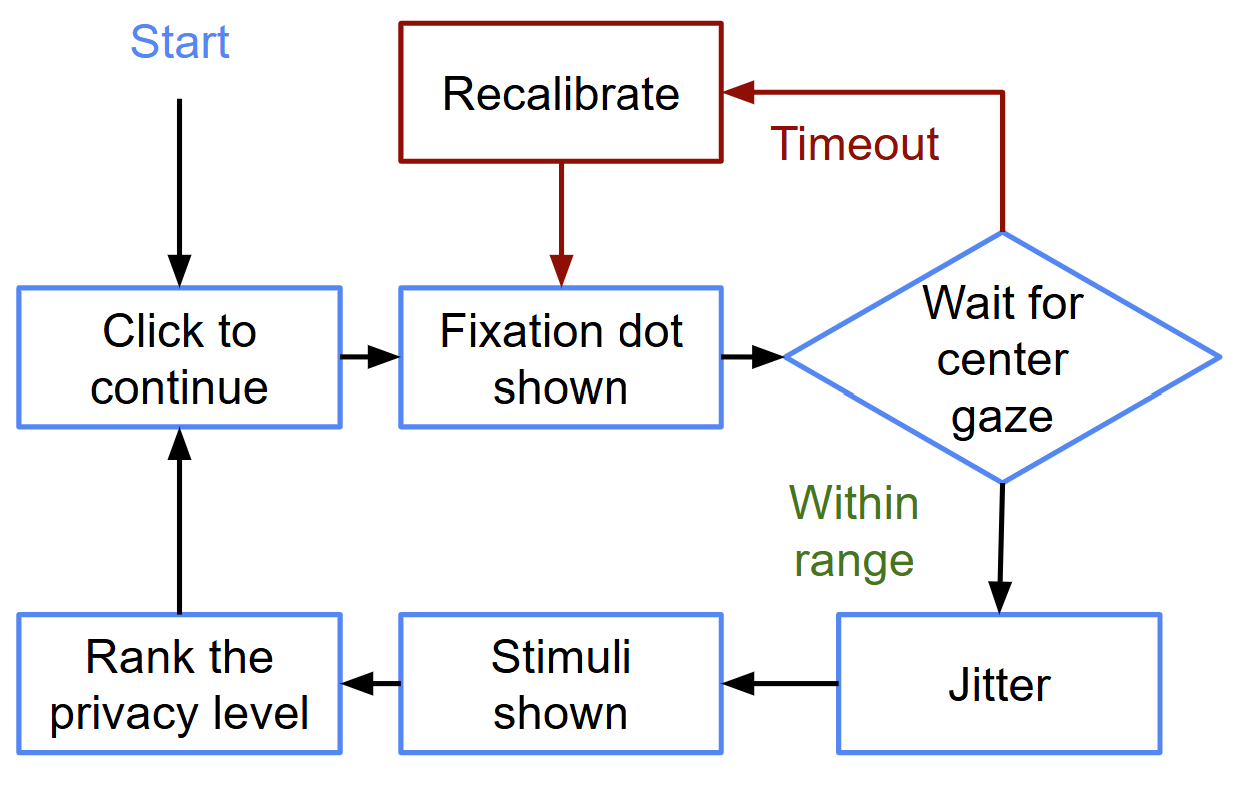}
  \caption{A sample trial flow-chart}
  \label{fig:flow}
\end{figure}

A sample trial is conducted as shown in Figure \ref{fig:flow}.

The stimuli were gathered from the validation and test sets of the VISPR dataset \cite{orekondy2017advisor}. 
Every block contained writing-based and human stimuli to avoid bias, with a distribution of private attributes as shown in Table \ref{tab:blocks}.

\begin{table}[h]
  \caption{Stimuli distribution over blocks}
  \label{tab:blocks}
  \begin{tabularx}{0.5\textwidth}{lXX}
    \toprule
    \textbf{Block} & \textbf{Attribute} & \textbf{Category}\\
    \midrule
    1 & Fingerprint, Receipts, Occupation, Sexual Orientation, Political Opinion &  Personal Description, Documents, Employment, Personal Life, Personal Life\\
    \midrule
    2 & Signature, Tickets, Medical Treatment, Personal Occasion, Home address &  Personal Information, Documents, Health, Personal Life, Whereabouts \\
    \midrule
    %3 & Race, Passports, Online Conversations, Social Circle, Phone Number & Personal Information, Documents, Internet, Relationship, Personal Life  \\
    3 & Face Complete, Credit Card, Medical History, Email Content, Religion &  Personal Description, Documents, Health, Internet, Personal Life \\
    \midrule
    4 & Full Name, Mail, License Plate Complete, Personal Relationship, Visited Location &  Personal Information, Documents, Automobile, Personal Life, Whereabouts \\
  \bottomrule
\end{tabularx}
\end{table}

\subsection{Data Collection}\label{app:data-collection}
To accurately track participants’ gaze and record eye movement data during the experiment, we first conduct a 13-point calibration to ensure the eye tracker is accurately mapping gaze coordinates to the screen, followed by a validation check to confirm that gaze accuracy is within acceptable error limits (e.g., <0.5°). We further recalibrate if drift or errors occur during the session.
The stimuli are then presented, and data streams are recorded.

%\linelabel{q12:1}
%\added[id=mayar]{Essential methodological safeguards were implemented to maintain participant engagement and ensure data quality throughout the experiment. These included randomised attention checks using stimuli with clearly private or non-private attributes to verify attentiveness, as well as structured breaks and interface resets to reduce fatigue. Moreover, the order of tasks and stimuli was randomised across participants to control for order and learning effects. These procedures follow established best practices in behavioural and user research.}
Essential methodological safeguards were implemented to maintain participant engagement and ensure high data quality throughout the experiment. These safeguards served to minimise bias, reduce participant fatigue, and confirm that responses reflected genuine attention and comprehension.
\\
Randomised attention checks were embedded at multiple points during the study. These checks consisted of stimuli with clearly identifiable attributes—some explicitly private (e.g., credit cards) and others non-private (e.g., randomly generated colour patches). Participants were asked to classify or respond to these items. Correct responses indicated attentiveness, while incorrect answers were flagged for potential disengagement or misunderstanding.
\\
To mitigate cognitive fatigue, structured breaks were introduced at predefined intervals (i.e. between tasks), allowing participants to rest and maintain focus. Participants were also allowed to stop the experiment at any point, if needed.
\\
To address order and learning effects, both the sequence of experimental tasks and the presentation order of stimuli were fully randomised for each participant. This ensured that performance patterns could not be attributed to predictable task progression, practice effects, or fatigue tied to task order.
\\
All these procedures align with established best practices in behavioural and user research \cite{carter2020best, hessels2025fundamentals, hooge2025fundamentals, nystrom2025fundamentals, niehorster2025fundamentals}, ensuring internal validity while safeguarding the participant experience.

\subsection{Data Processing}
We cleaned and processed raw eye-tracking data for further analysis.
The raw data is extracted from the EyeLink Data Viewer as .edf files, including the timestamped gaze coordinates (X, Y), fixation and saccade metrics (e.g., duration, amplitude, velocity), and pupil size. We then used the stimulus metadata to map gaze coordinates to specific ROIs, categorised gaze events into relevant regions for analysis, and merged eye-tracking data with task-specific inputs (e.g., participant ratings and mouse clicks).

We structured the processed data into a usable dataset for analysis by organizing data into rows and columns ( .csv format), including information such as participant demographics, stimuli details, and task condition.
Finally, we formatted the data for statistical or machine learning tools (e.g., Python) in the following structure:

\begin{forest}
for tree={
  grow'=0,                      % grow tree horizontally
  child anchor=west,           % anchor children on the west (left)
  parent anchor=east,          % anchor parent on the east (right)
  anchor=west,
  edge={->, >=latex},          % draw arrows
  l sep=20pt,                  % level separation
  s sep=15pt,                  % sibling separation
  font=\sffamily,              % optional: sans serif font
}
[Block n
  [Participant1
    [Task1
      [Stimulus1]
      [Gaze1]
      [Rating1]
      [...]
    ]
    [Task2
      [Stimulus1]
      [Gaze1]
      [Rating1]
      [...]
    ]
  ]
  [...]
  [Participant100
    [Task1
      [...]
    ]
    [Task2
      [...]
    ]
  ]
]
\end{forest}

\section{ML Implementation Details}\label{apx:implementation}

\added[id=mayar]{Given a dataset of inputs (e.g. gaze data) annotated with labels (e.g. user-selected privacy levels), we define the training set as:}
$$\mathcal{D} = \{(\mathbf{x}_i, y_i)\}_{i=1}^N$$
\added[id=mayar]{where: $\mathbf{x}_i \in \mathbb{R}^d$  represents the gaze data of the $i^\text{th}$ sample (or a feature vector extracted from the stimuli and the gaze when both are inputs, as defined by each task),  $y_i \in \mathcal{Y} = \{1, 2, \dots, K\}$  is the corresponding label (e.g. user-provided privacy level or attribute name or privacy expertise, as defined by each task).}

\paragraph{The training phase:}
In \added[id=mayar]{a supervised learning setting like classification, the training phase uses labelled data—pairs of input features ($x_i$) and their corresponding ground truth labels ($y_i$). During training, the model learns a mapping from inputs to labels by minimising a loss function that measures the discrepancy between predicted and true labels.}

\added[id=mayar]{The goal is to learn a function  $f_\theta: \mathbb{R}^d \rightarrow \mathcal{Y}$, parameterized by $\theta$, that minimizes the classification loss:}
$$\mathcal{L}(\theta) = \frac{1}{N} \sum_{i=1}^N \ell(f_\theta(\mathbf{x}_i), y_i)$$
\added[id=mayar]{where $\ell(\cdot, \cdot)$ is the loss function.}

\paragraph{The inference phase:}
During \added[id=mayar]{the inference (testing) phase, the model receives only input features from new, unseen samples and generates predicted labels based on the patterns it learned during training. No true labels are available during inference; predictions are made autonomously using the trained model parameters.}

\added[id=mayar]{More technically, a new unseen gaze sample $\mathbf{x}^\ast \in \mathbb{R}^d$ is processed through the trained model to produce a predicted label:}

$$\hat{y} = f_{\theta^\ast}(\mathbf{x}^\ast)$$

\added[id=mayar]{We then compare the predictions $\hat{y}_i$ with the true labels $y_i$ only to quantify the models' performance.}

\paragraph{Train-test split:}
We \added[id=mayar]{employ cross-validation to optimise generalisation performance and prevent overfitting. In cross-validation, the data is split into 
$k$ folds; the model is trained on 
$k-1$ folds and tested on the remaining one, repeating this process $k$ times so that each fold serves as a test set once. The final performance is averaged across all folds, providing a more robust and unbiased estimate than a single train-test split. Here, we used an 80-20 train-test split.}

\paragraph{ML models:}
In \added[id=mayar]{our implementation, we mostly employed classical machine learning algorithms available within the \textit{Scikit-learn} framework, using their default configurations. \textit{Decision Trees} were implemented using 'DecisionTreeClassifier'. \textit{Support Vector Machines (SVMs)} were used via 'SVC' with an 'RBF kernel'. \textit{Logistic Regression} was applied using 'LogisticRegression' with L2 regularisation. For \textit{Random Forests}, we utilised 'RandomForestClassifier', an ensemble of decision trees trained on bootstrapped subsets of data with feature bagging. \textit{K-Nearest Neighbors (KNN)} was implemented using 'KNeighborsClassifier', which assigns class labels based on the majority vote among the k most similar training samples in feature space.
Further information and default hyperparameters can be found in the Scikit-Learn documentation:} \url{scikit-learn.org/stable/supervised_learning.html}. 
\added[id=mayar]{Finally, for other advanced models, we used the original implementations provided by the referenced papers.}

\section{Potential Real-World Applications}\label{apx:apps}
\added[id=mayar]{In this section, we discuss the potential SPP and UPP applications in more details:}

\bpara{SPP applications} \added[id=mayar]{Stimuli-based perceived privacy results can be used in applications such as:}
\begin{itemize}[leftmargin = *]
    \item \textbf{Setting hand-picked parameters in privacy-preserving protocols}: \added[id=mayar]{Similar to our DP application in \cref{sec:applications}, federated learning (FL) \cite{mcmahan2017communication, kairouz2021advances} can also benefit from our approach. For example, the \textit{model update perturbation} step involves adding random noise to the model updates (usually gradients or weights) before sending them to the central server. This includes noise scale $\sigma$ and a clipping norm $c$ parameters. These are typically hand-picked through empirical tuning or heuristics to achieve a privacy-utility tradeoff. Hence, \methodName~can be used to implicitly set these parameters and personalise them to the users' expectations. The same applies to K-anonymity \cite{samarati1998protecting}, L-diversity \cite{machanavajjhala2007diversity}, T-closeness \cite{li2006t}, to set the $k$, $L$, and $T$ values.}

    \item \textbf{Privacy auctions:} \added[id=mayar]{Privacy auctions \cite{ghosh2011selling, zhang2020selling} are mechanisms where users "sell" their private data or privacy loss in exchange for compensation. These auctions aim to determine how much privacy loss  users are willing to tolerate, and at what cost, allowing systems to personalise privacy levels across individuals based on their preferences.  Users specify a subjective cost or price that they associate with a unit of privacy loss, typically through surveys.
    Hence, \methodName~offers a more robust and implicit feedback with the need for users to understand and evaluate their data.}

    \item \textbf{Synthetic data generation:} \added[id=mayar]{Synthetic data generation \cite{liu2022privacy} refers to the creation of artificial data that mimics the statistical properties of real user data, often used to preserve privacy while enabling data analysis or model training. Here, again \methodName~can be used to match the user privacy expectations with respect to the data, e.g. data fidelity preferences (acceptable levels of distortion or allowed deviation from real data distributions).}

    \item \textbf{Access control models:}\added[id=mayar]{
    Access control models regulate who can access what data under which conditions. Two key types are: (i) Attribute-Based Access Control (ABAC) \cite{hu2015attribute}, which dynamically grants or denies access based on user, resource, and environmental attributes (e.g., role, location, time). The stimuli-specific insights of \methodName,  can be used to automatically infer these attributes (e.g. \textit{the private attribute recognition task}) and set the corresponding privacy parameters (e.g. access) without explicitly defining and listing all possible roles (e.g. \textit{the privacy expertise} or the \textit{user profiling} tasks).
    (ii) Human-in-the-loop privacy controls, which empower users to manage privacy interactively in real-time systems. In this case, \methodName can be integrated with, for example, Instagram’s 'Restrict' feature \cite{parmelee2020insta} or e-mail spam filters to allow users to flag content without explicitly selecting buttons and going through lists.}  
\end{itemize}

\bpara{UPP applications} \added[id=mayar]{User-specific insights can also be integrated into several real-world applications such as:}

\begin{itemize}[leftmargin = *]
    \item \textbf{Privacy nudges:} \added[id=mayar]{Privacy nudges \cite{ioannou2021privacy} are subtle interventions designed to guide users toward making more privacy-conscious decisions without restricting their freedom of choice.
    They are commonly used in social media platforms, e.g. Facebook’s contextual privacy warnings \cite{meta_sensitive_content_2024} and Chrome browser-based security warnings \cite{google_safebrowsing}
    These nudges typically provide a uniform textual content—such as generic privacy explanations—to all users, regardless of individual differences in privacy literacy. 
    As a result, they risk producing mismatched comprehension: users with limited privacy knowledge may struggle to understand the information (under-comprehension), while more knowledgeable users may find it redundant or oversimplified (over-comprehension), ultimately reducing the effectiveness of the intervention.
    Hence, they can be personalised based on a user's expertise or profile, implicitly via \methodName.}
    \item \textbf{Privacy labels and transparency notices:} \added[id=mayar]{Similar to privacy nudges, \methodName~can be used in privacy labels and transparency notices, e.g. Apple’s App Store privacy labels \cite{apple_privacy_labels}, to help users understand and control their privacy choices according to their profiles to avoid mismatched comprehension.}
    \item \textbf{Cohort-based recommendations:} \added[id=mayar]{Cohort-based recommendations, e.g. Google’s Federated Learning of Cohorts (FLoC) \cite{google_floc_whitepaper}, group users into segments (or cohorts) based on shared characteristics, such as behaviour, preferences, or demographics, and generate recommendations tailored to each group. Instead of personalising for individuals, the system provides suggestions optimised for the typical member of a cohort, balancing personalisation and privacy by avoiding the need for fine-grained individual profiling. 
    Here, again, \methodName~can be used to profile users according to their gaze behaviour such as privacy preferences.}
\end{itemize}

\section{Cognitive Theories and Privacy}\label{apx:theories}
Here, we give more details about some cognitive theories that further support our mapping functions.

\paragraph{The Exponential Mapping} captures a perception that escalates rapidly with small increases in privacy risk or sensitivity.
The related theories include:
\newline
(i) The prospect theory \cite{kahneman2013prospect, choi2014prospect} where people weigh potential losses and gains and can provide a direct assessment of privacy levels based on their perception of risks and benefits. Individuals may perceive increasing risk with diminishing marginal tolerance—well captured by an exponential curve. The exponential shape models the non-linear, often risk-averse valuation of privacy losses.
\newline
(ii) The risk-reward trade-off theory \cite{park2012affect} where individuals balance risks and rewards in a unified decision, leading to a cumulative rating of privacy. 
This model implies that perception accumulates as users weigh these aspects, with increasingly steep aversion to risk—supporting an exponential model for mapping ratings to privacy budgets.
\newline
(iii) The communication privacy management (CPM) theory \cite{petronio2021communication} frames privacy as the control of boundary permeability based on accumulated context and sensitivity. The control intensifies sharply as users assess perceived violations, again suggesting an exponential increase in perceived privacy sensitivity.
\newline
(iv) The theory of planned behaviour \cite{saeri2014predicting} where attitudes, norms, and perceived control influence a single privacy decision, often resulting in a direct rating.
The interaction of these variables can collectively lead to a compounded privacy concern that builds up non-linearly, fitting an exponential growth in privacy valuation.

\paragraph{The Sequential Mapping} corresponds to decision-making that unfolds in discrete steps. The following theories underpin this logic:
\newline
(i) Contextual integrity \cite{nissenbaum2004privacy} sequentially considers factors like actors, attributes, and transmission principles. For example, a user might first decide if a context violates norms, then determine the severity of the violation, resulting in a layered decision path consistent with step-wise or rule-based mappings. 
\newline
(ii) The privacy paradox \cite{kokolakis2017privacy} suggests that users might decide in one step whether to share information and then, based on cognitive dissonance, adjust how much information they disclose or rate its privacy (i.e. a post-hoc justification). This aligns with a sequential structure where decisions are refined over time.
\newline
(iii) The dual-process theory \cite{aivazpour2017unpacking} assumes that automatic, heuristic-driven decisions (e.g., 'private or not') may precede more deliberative evaluations of privacy levels.
\newline
(iv) The heuristic-systematic model \cite{gharib2024towards} suggests that a heuristic (i.e., quick judgment) may guide the first decision, followed by a deeper systematic analysis to refine privacy preferences.

\paragraph{The Sigmoid Mapping}
reflects bounded sensitivity at both extremes: users are easily decide on very safe or very private data, but become highly sensitive in an intermediate uncertainty zone.
The following theories justify this mapping:
\newline
(i) The bounded rationality theory \cite{simon1990bounded} suggests that decisions are made with limited information, often leading to “good enough” rather than systematically cumulative or sequential outcomes. Users might skip steps or make fragmented decisions. The sigmoid captures this minimal sensitivity at low and high certainty, with steep reactivity in ambiguous cases.
\newline
% \mayar{recheck:}
% (ii) The cultural dimensions theory \cite{smith2020cultural} argues that decisions depend on cultural factors (e.g., individualism vs. collectivism), which influence the overall approach to privacy without necessarily being cumulative or sequential. 
(ii) The uncertainty reduction theory \cite{knobloch2008uncertainty} suggests that decisions aim to reduce uncertainty and may involve multiple rounds of information gathering and refinement.
% (iii) The attribution theory \cite{manusov2008attribution} claims that decisions depend on judgments of others' intentions, which may be dynamic and context-specific, leading to unstructured privacy assessments. 
\newline
(iii) The cognitive dissonance theory \cite{harmon2012cognitive} suggests that users adjust decisions retroactively to reduce dissonance. This retroactive calibration results in smooth but non-linear adjustments over time—reflected in the sigmoid's gentle asymptotes and steep central slope.

\section{Adverserial Perspective on Differential Privacy}\label{sec:appendix-dp}
\newcommand{\mech}{M}
We want to briefly motivate the definition of the adversarial advantage for the differential private mechanism we use in \cref{sec:preliminaries} and \cref{sec:applications}.

We use the following security game for a DP-mechanism $\mech$, an adversary $\mathcal A$ and a challenger $\mathcal C$.

\begin{enumerate}    
    \item The adversary $\mathcal A$ chooses two valid adjacent inputs sets $D_0,D_1$ for $\mech$ and sends them to the challenger.
    \item The challenger samples a bit $b$. It runs $\mech(D_b)$ a random number of times and stores the outputs in a set $S$. 
    \item Upon receiving $S$, $\mathcal A$ outputs a bit $b'$.
\end{enumerate}
The adversary wins the security game if $b'=b$.
The advantage of $\mathcal A$ is defined as $\operatorname{adv}_{\mathcal A}\coloneqq 2p-1$, where $p$ is the maximal probability that $\mathcal A$ wins for \emph{any} $S$. Note that the adversary in this game is exceptionally strong since it only needs to win the game for one specific output set $S$ (no matter how unlikely $S$ itself is).
The setup is nevertheless relevant, since in real-world use cases, unlikely outputs might nevertheless occur, and even then, the privacy of the input data should be preserved.

In a slight abuse of notation, we also denote by $\operatorname{adv}_{\mathcal A}$ the maximal advantage achieved by a ppt. adversary $\mathcal A$. We want to determine an upper bound on the advantage.
Given the limited information available the most successful adversary $\mathcal A$ outputs $b'=0$ if %\linelabel{q6:2}
\added[id=PR]{$\Pr(b=0|S)\coloneqq \Pr(D_0|M(D_0)\in S)\geq \Pr(D_1|M(D_1)\in S)\eqqcolon \Pr(b=1|S)$} and $b'=1$ otherwise.
Let w.l.o.g. $\Pr(b=0|S)\geq \Pr(b=1|S)$.
Hence, this adversary has an advantage
$\operatorname{adv}_{\mathcal A}=2\Pr(b=0|S)-1$.
If the mechanism satisfies $\varepsilon$-differential privacy, \cref{eq:dp} implies $\Pr(S|b=0)\leq e^{\varepsilon}\Pr(S|b=1)$.
But $\Pr(S|b=i)=\frac{\Pr(S,b=i)}{\Pr(b=i)}=2\Pr(S,b=i)=2\Pr(S)\Pr(b=i|S)$ for $i=0,1$.
Thus $\Pr(S|b=0)\leq e^{\varepsilon}\Pr(S|b=1)\Rightarrow 
\Pr(b=0|S)\leq e^{\varepsilon}\Pr(b=1|S)=e^{\varepsilon}(1-\Pr(b=0|S))\Rightarrow \Pr(b=0|S)\leq \frac{e^{\varepsilon}}{1+e^{\varepsilon}}$.
We conclude that $\operatorname{adv}_{\mathcal A}=2\Pr(0|S)-1\leq \frac{e^{\varepsilon}}{1+e^{\varepsilon}}-1=\frac{e^{\varepsilon}-1}{1+e^{\varepsilon}}$.